\documentclass[fleqn,usenatbib]{mnras}

\usepackage[T1]{fontenc}
\usepackage{ae,aecompl}

\usepackage{graphicx}	
\usepackage{amsmath}	
\usepackage{amssymb}	
\usepackage{siunitx}
\usepackage{amsbsy}
\usepackage{xcolor, wasysym, ulem}
\usepackage{float}
\usepackage{caption}
\usepackage{subcaption}
\usepackage{enumerate}
\usepackage{lscape}
\usepackage{rotating}
\usepackage{hvfloat}
\usepackage{newtxtext,newtxmath}

\newcommand{\whz}{\,{\rm W\,Hz^{-1}}}
\newcommand{\Lum}{{L_{\rm 1.4\,GHz}}}
\newcommand{\lum}{{L_{\rm 400\,MHz}}}
\newcommand{\bra}[1]{\langle #1\rangle}
\newcommand{\qmir}{q_{\rm 24\upmu m}}
\newcommand{\qfir}{q_{\rm 70\upmu m}}
\newcommand{\qtir}{q_{\rm TIR}}
\newcommand{\qtirbol}{q^{\rm RC}_{\rm TIR}}
\newcommand{\tdust}{T_{\rm dust}}
\newcommand{\lmir}{L_{\rm 24\upmu m}}
\newcommand{\lfir}{L_{\rm 70\upmu m}}
\newcommand{\ltir}{L_{\rm TIR}}
\newcommand{\smir}{\sigma_{\rm 24\upmu m}}
\newcommand{\sfir}{\sigma_{\rm 70\upmu m}}
\newcommand{\stir}{\sigma_{\rm TIR}}
\newcommand{\lrc}{L_{\rm RC}}

\title[Faint sources in ELAIS-N1]{{Deep uGMRT observations of the ELAIS-North\,1 field: statistical properties of radio--infrared relations up to $z \sim 2$}}

\author[Akriti Sinha et al.]{Akriti Sinha,$^1$\thanks{E-mail: \href{mailto:sinha.akriti44@gmail.com}{sinha.akriti44@gmail.com} (AS); {\href{mailto:abasu@tls-tautenburg.de}{abasu@tls-tautenburg.de} (AB)}}
Aritra Basu,$^{2\,\star}$ Abhirup Datta$^1$ and Arnab Chakraborty$^{1,3}$ 
\\
$^{1}$Department of Astronomy, Astrophysics and Space Engineering, Indian Institute of Technology, Indore 453552, India\\
$^2$Th\"{u}ringer Landessternwarte, Sternwarte 5, 07778 Tautenburg, Germany\\
$^3$Department of Physics and McGill Space Institute, McGill University, Montreal, QC, Canada H3A 2T8\\}

\date{Last updated 2015 May 22; in original form 2013 September 5}

\pubyear{2020}

\begin{document}
\label{firstpage}
\pagerange{\pageref{firstpage}--\pageref{lastpage}}
\maketitle

\begin{abstract}

Comprehending the radio--infrared (IR) relations of the faint extragalactic radio sources is important for using radio emission as a tracer of star-formation in high redshift ($z$) star-forming galaxies (SFGs). 
Using deep uGMRT observations of the ELAIS-N1 field in the 0.3--0.5\,GHz range, we study the statistical properties of the radio--IR relations and the variation of the `$q$-parameter' up to $z=2$ after broadly classifying the faint sources as SFGs and AGN.
We find the dust temperature ($\tdust$) to increase with $z$. This gives rise to $\qmir$, measured at $24\,\upmu$m, to increase with $z$ as the peak of IR emission shifts towards shorter wavelengths, resulting in the largest scatter among different measures of $q$-parameters. $\qfir$ measured at $70\,\upmu$m, and $q_{\rm TIR}$ using total-IR (TIR) emission are largely unaffected by $\tdust$.
We observe strong, non-linear correlations between the radio luminosities at 0.4 and 1.4\,GHz with $70\,\upmu$m luminosity and TIR luminosity($\ltir$). To assess the possible role of the radio-continuum spectrum in making the relations non-linear, for the first time we study them at high $z$ using integrated radio luminosity ($\lrc$) in the range 0.1--2\,GHz.
In SFGs, the $\lrc$--$\ltir$ relation remains non-linear with a slope of $1.07\pm0.02$, has a factor of 2 lower scatter compared to monochromatic radio luminosities, and $\qtirbol$ decreases with $z$ as $\qtirbol = (2.27 \pm 0.03)\,(1+z)^{-0.12 \pm 0.03}$. A redshift variation of $q$ is a natural consequence of non-linearity. We suggest that a redshift evolution of magnetic field strengths and/or cosmic ray acceleration efficiency in high-$z$ SFGs could give rise to non-linear radio--IR relations.

\end{abstract}

\begin{keywords}

radio continuum: galaxies, infrared: galaxies, galaxies: active, galaxies: ISM
\end{keywords}

\section{Introduction}

Deep radio-continuum observations at micro-Jansky ($\upmu$Jy) level below about 10\,GHz provide an unobscured view of the extragalactic sky up to a very early Universe \citep[e.g.,][]{Condon_1992, Williams_2016, Novak_2017}. Of late, deep observations have enabled the study of the properties of a diverse population of sources, ranging from star-forming galaxies (SFGs), typically at the faint flux density end, through radio-quiet and Fanaroff–Riley (FR) class 0-type active galactic nuclei (AGN) with intermediate flux densities (few tens of mJy), to powerful FRI- and FRII-type AGN at the high flux density regime ($\gtrsim100$\,mJy). These observations not only facilitate finding and determining the relative abundance of these sources, but when combined with multi-waveband information, they provide an excellent means to statistically study their intrinsic properties, impact of their environment, and constrain their evolution over the history of the Universe \citep[see, e.g.,][]{Padovani_2009, Padovani_2011, Padovani_2015, Novak_2018, Baldi_2019, Tisanic_2019, Mingo_2019, Hardcastle_2019, Hardcastle&Croston_2020}. 

Broadly speaking, in terms of the origin of emission, extragalactic sources can be classified into AGN and SFGs. Although the emission mechanism in both classes of sources is dominated by synchrotron radiation, AGN are powered by relativistic jets launched by the central black hole, while for SFGs, the emission originates from cosmic ray electrons (CREs) accelerated in the shock fronts of supernovae explosions. 
Deep radio-continuum surveys with sensitivity $\lesssim100\,\upmu$Jy are opening up a new window on what is usually considered the `normal' galaxy population at high redshift ($z$). 
Since radio observations are free from dust obscuration, unlike ultraviolet (UV) and H$\alpha$ tracers, and the source-photometry are generally not confused, unlike mid- to far-infrared tracers \citep[e.g.,][]{Madau_2014, Jarvis_2015}; radio continuum emission is being used to trace the star-formation history of the Universe and the evolution in the star forming main sequence \citep{Daddi_2007, Seymour_2008, Novak_2017, Ocran2020, Leslie_2020}. Lately, the scenario of co-evolution is emerging, wherein feedback from star formation and jet launching play an important role in the evolution of the AGN and SFG populations \citep{Jurlin_2020, Webster_2021}. Therefore, it is crucial to carefully characterize the properties of the sources in order to determine the dominant component, i.e., star-formation or the AGN activity, which is contributing to the radio continuum emission. This in turn, enables a contamination-free estimation of the star formation rate (SFR) from radio continuum measurement alone.

Using radio continuum emission as a tracer of SFR relies on one of the tightest, near-linear correlations in astrophysics observed between the radio and infrared (IR) luminosities, the \textit{radio--IR} relation \citep[e.g.,][]{Helou_1985, Condon_1992, Yun_2001}. This relation spans over five orders of magnitude in luminosity with dispersion less than a factor of two, and holds good from dwarf \citep[][]{chyzy11, roych12, jurus14} to ultra-luminous infrared galaxies (ULIRGs) on galaxy-integrated scales \citep[e.g.,][]{Appleton_2004, Sargent_2010_1, Mao_2011, Basu_2015}. It is believed that the UV photons from massive ($\gtrsim10\,\rm M_\odot$) OB-type stars, that are absorbed by the dust, is re-radiated in the infrared wavebands and the same OB-type stars provide the synchrotron emitting CREs when they end their short lives (up to a few Myr) as supernovae explosions, giving rise to the correlation.

However, a number of seemingly independent physical parameters of the interstellar medium (ISM) are responsible for the emission processes in the radio and infrared wavebands, such as, the number density of CREs; energy losses and escape of CREs; magnetic field amplification mechanism; star formation history; dust absorption efficiency; and densities of dust and gas. The tightness and the slope of the radio--IR relation depends on the interplay between these physical parameters, and on whether or not energy equipartition between magnetic fields and CREs are valid \citep{Volk_1989, Helou_bicay1993, Niklas_Beck_1997, Bell_2003, Lacki_2010, berkhuijsen2013, Basu_2017}. Based on theoretical and empirical results, a framework based on efficient amplification of the magnetic fields via supernovae-driven fluctuation dynamo in galaxies has been put forward that connects magnetic field strengths and cosmic ray, and gas densities, to explain the tightness and cosmic evolution of the radio--IR relation \citep[see, e.g.,][]{Lacki_Thompson_2010, Schleicher_Beck_2013, Schober_2016}. These studies bring to light that the relation is expected to evolve with redshift due to a combination of evolution of the properties of the ISM, or confinement of CREs by the magnetic fields, manifesting as either a change in the slope, making it significantly non-linear, or a change in the ratio of infrared to radio luminosities. Therefore, it is essential to study the properties of radio--IR relations at high redshifts.

Radio continuum emission in star-forming galaxies mostly originates from the synchrotron and the free--free mechanisms. Owing to the steep spectrum of the synchrotron emission, it dominates at frequencies $\lesssim 2$\,GHz \citep[][]{basu12a}. Since the free--free emission with flat spectrum directly originates as a consequence of star formation, high frequency ($\gtrsim 20$\,GHz) radio observations are well suited to constrain the cosmic star formation history \citep{Murphy_2011, murph15}. However, performing deep, large sky-area surveys at these frequencies is time expensive due to the relatively small field-of-view, and the emission is contaminated by anomalous microwave emission \citep{Leitch_1997, murph10}.
Low frequency radio surveys significantly below 1\,GHz are important as the rest-frame emission is dominated by the synchrotron emission, and are relatively less contaminated by the free--free emission as compared to observations near or above 1\,GHz. Furthermore, the emission from the AGN is typically optically thick and less variable at frequencies below 1\,GHz \citep{Condon_2016}. This give rise to less biases in radio photometry when compared to high radio frequency observations, making low-frequency observations well suited for identifying steady AGN emission and thereby studying the star formation history via the radio--IR relations.

Ongoing and future sensitive large sky-area radio continuum surveys, such as, the LoTSS using the LOFAR \citep{Shimwell_2017}, the VLASS using the Karl G. Jansky Very Large Array \citep[VLA;][]{Lacy_2020}, 
and the MIGHTEE using the MeerKAT \citep{Jarvis_2016}, and surveys using the Square Kilometre Array \citep[SKA;][]{Jarvis_2015SK} and next-generation VLA \citep[ngVLA;][]{Francesco_2019} later in the decade, are going to detect several tens of millions of radio sources. Robust characterization of all the sources, primarily based on optical and/or infrared spectroscopy, for example, using the James Webb Space Telescope \citep[JWST;][]{Kalirai_2018} is going to be a challenging proposition, and much of the initial source classification is expected to rely on existing, ancillary multi-waveband data. Therefore, it is crucial to investigate the efficacy of source classification based on existing photometric surveys in the optical and infrared wavebands, and using relatively shallow but large sky-area spectroscopy from the Sloan Digital Sky Survey (SDSS) data and investigate their impact on the radio--IR relations.

In order to prepare for these large sky-area surveys, it is important to first investigate smaller sky-areas which are prototypical examples of future surveys. To this end, we have performed deep observations of the European Large Area ISO Survey--North\,1 (ELAIS-N1) field covering an area of 1.8\,deg$^2$ with a root mean square (rms) sensitivity of $\sim 15\,\rm \upmu Jy\,beam^{-1}$ centered at 400\,MHz using the upgraded Giant Metrewave Radio Telescope (uGMRT) presented in \citet{Arnab2019_2}. In this paper, we broadly classify the radio sources into AGN and SFGs using publicly available ancillary multi-waveband data to investigate the radio--IR relations up to $z\sim2$. This paper is organised as follows: Section~\ref{sec_data} describes the radio and multi-wavelength data used in this work. The different methods of identifying SFGs and AGN are described in Section~\ref{sec:agn_sfg}. The radio and infrared spectral energy distribution (SED) fitting for $k$-correcting to the rest-frame are described in Section~\ref{sec:radio_IR_SED}. In Section~\ref{sec:result}, we present our results on the statistical properties of the radio--IR relations, and discuss them in Section~\ref{sec:discussion}. A summary of our work is presented in Section~\ref{sec:summary}. In this work, we have used the best-fit cosmological parameters from the Planck\,2018 results \citep{Planck_2020}: $\Omega_\Lambda = 0.68$, $\Omega_{\rm m} = 0.31$ and $H_0 = 67.36\,\rm km\,s^{-1}\,Mpc^{-1}$.

\section{Data}\label{sec_data}

In this section, we present our observations of the ELAIS-N1 field using the uGMRT between 300--500\,MHz, and discuss the salient features of radio continuum data at other radio frequencies. In addition, we also discuss in brief the assorted multi-wavelength survey data used for further analyses in this paper. The salient features of the different surveys used in this work are summarized in Table~\ref{tab:data}. In this work, we have identified the counterparts of our uGMRT sample by cross-matching them to their nearest neighbour in various multi-wavelength catalogues. We have used a search radius of 3\,arcsec for all datasets except at 1.4\,GHz where a search radius of 5\,arcsec was used.

\begin{table*}
    \centering
    \caption{Salient features of various multi-waveband surveys of the ELAIS-N1 field. The columns represent the multi-wavelength catalogues with total area covered, resolution, and, corresponding $5\sigma$ sensitivity in mJy. The column `Size' represents the number of sources in a catalogue with 400-MHz uGMRT counterpart, and their corresponding percentage are listed in the last column.}
    \begin{tabular}{cccccc}
    \hline 
    Catalogue & Total Area  &   Resolution & $5\sigma$ sensitivity   & Size & Percentage \\
    &  (deg$^2$) &   ($''\, \times \, ''$) & (mJy) & & 
            \\[1ex]
    \hline   
    uGMRT 400 MHz & 1.8 & $4.6 \times 4.3$ & 0.075  &  2528$^\dagger$  &   100 \\
    LoTSS & 64 & $6 \times 6$ & 0.1  & 2225 & 88\\ 
    GMRT 612 MHz & 1.13 & $6 \times 6$  & 0.04 &  1518 &   60 \\
    FIRST  &  Large area survey  & $5 \times 5$ & 0.75    &  144  &   6 \\
    BOSS   & Large area survey    &   &   &  597  & 24 \\
    SWIRE all IRAC bands & 2.0 & &  &  1470   &  58 \\
    SWIRE 24 $\upmu$m & 8.72  & $5.6 \times 5.6$ & 0.45  & 1201  &  48\\
    SWIRE 70 $\upmu$m & 8.72  & $16.7 \times 16.7$ & 2.75  & 388   &  15\\
    HerMES 250 $\upmu$m  & 3.25 & $18.2 \times 18.2$ & 25.8 &  702  &  28    \\
    HerMES 350 $\upmu$m  & 3.25 & $25 \times 25$ & 21.2 & 686   &    27   \\
    HerMES 500 $\upmu$m  & 3.25 & $36.3 \times 36.3$ & 30.8 & 557   &    22  \\
    Redshifts &  &  &    &  2319 & 92 \\
    \hline
    \end{tabular}
    \begin{flushleft}
    $^\dagger$ Represents the number of sources compiled above $\sim 6\,\sigma$ with point source sensitivity $\gtrsim 100\,\upmu$Jy. All other catalogues are matched to these 2528 sources. 
    \end{flushleft}
    \label{tab:data}
\end{table*}

\subsection{Radio continuum data}

\subsubsection{uGMRT observations at 400 MHz}
Observations of the ELAIS-N1 field, centered at $\rm RA=16h\, 10m\, 1s, Dec=54d\, 30m\, 36s$ (J2000), were carried out in May--June 2017 using the uGMRT for a total of 25\,hrs (including calibration overheads) spanning over four nights (proposal code: 32\_120). These observations were performed in Band\,3 covering the frequency range 300 to 500\,MHz centered at 400\,MHz using the new GMRT wideband (GWB) correlator with a frequency resolution of 24\,kHz. A total of $\approx13$\,hrs were spent on the target field. The final image of the ELAIS-N1 field is obtained at an angular resolution of $4.6\arcsec \times 4.3\arcsec$ with an rms noise of $15\,\upmu$Jy\,beam$^{-1}$, covering a $1.8$\,deg$^2$ region. A catalogue comprising of total 2528 sources above $6\,\sigma$ with point source sensitivity $\gtrsim 100\,\upmu$Jy was generated using P{\tiny Y}BDSF  \citep{Mohan_Raffrey2015} from these data. We refer an interested reader to \citet{Arnab2019_2} for a detailed description of the data analysis procedure and catalogue generation. For the purpose of our analysis, we have considered the total integrated flux densities at 400 MHz for these sources.

\subsubsection{LoTSS data at 146\,MHz}

The ELAIS-N1 field was observed as a part of the LOFAR Two-metre Sky Survey \citep[LoTSS;][]{Shimwell_2017,Shimwell_2019} observed using the Low Frequency Array (LOFAR). It is one of the deep fields of the LoTSS at 146.2\,MHz and reaches an rms noise of $20\,\upmu\rm Jy\,beam^{-1}$ in the central region at an angular resolution of $6\arcsec$. The publicly available LoTSS catalogue contains 84\,862 sources \citep{Sabater_2021} covering a sky area of 64\,deg$^2$, of which 16\,435 sources are detected within 1\,degree from the centre of the pointing. 2225 sources from the LoTSS catalogue were matched to our uGMRT 400-MHz catalogue.

\subsubsection{GMRT observations at 612 MHz}\label{sec:612MHz}

Observations of the ELAIS-N1 field were carried out at 612\,MHz using the legacy GMRT between 2011 and 2013 (project codes: 20\_044, 21\_083, 22\_056). These archival data were re-analyzed by \citet{Chakraborty_2020} covering an area of $\sim 1.13\,$deg$^2$ using seven pointings arranged in a hexagonal pattern centered on $\rm RA = 16{h}\,10{m}\,30{s}, DEC = 54d\,35m\,00s$ (J2000) \citep[see][]{Taylor&Jagannathan2016}. 
An rms of $\approx8\,\upmu$Jy\,beam$^{-1}$ at $6\arcsec$ resolution was obtained in the central region after mosaicking the seven pointings. Similar to the uGMRT data at 400\,MHz, the source catalogue at 612\,MHz was generated by applying P{\tiny Y}BDSF on the mosaicked map which provided a total of 2342 sources above a $6\,\sigma$-level, i.e., with point source sensitivity above $50\,\upmu$Jy \citep[see][for details]{Chakraborty_2020}.

\subsubsection{FIRST data at 1.4\,GHz}

We have also utilized the Faint Images of the Radio Sky at Twenty centimetres (FIRST) survey \citep{White_1997} that has used the Very Large Array to compile a catalogue of 946,432 sources covering a sky-area of 10,575\,deg$^2$ \citep{Helfand_2015}. FIRST covers the ELAIS-N1 region with a relatively shallow $5\sigma$ sensitivity limit of 0.75\,mJy at an angular resolution of $5\arcsec$. On cross-matching with our 400-MHz uGMRT catalogue, we find 144 sources that has also been detected in the FIRST survey.

\subsection{Ancillary multi-wavelength data}

In order to classify the detected radio sources as AGN and SFGs, and to perform $k$-correction, we have used publicly available data in the mid- to far-infrared (MIR and FIR, at 24 and 70\,$\upmu$m, respectively) wavelength regime, and optical spectroscopy from the Sloan Digital Sky Survey (SDSS). Here we discuss the salient features of these data.

\subsubsection{BOSS/SDSS Spectroscopy}

The Baryon Oscillation Spectroscopic Survey (BOSS) refers 
to the dark time survey of the third phase of the 
SDSS \citep[SDSS-III;][]{York_2000}. BOSS consists of two spectroscopic surveys over an area of 10,000 deg$^2$ \citep[see][for details]{Eisenstein2005,Eisenstein2011}. To observe ancillary science programs, a series of plates were added to the SDSS-III survey beyond 2012. Four plates were granted to observe and obtain spectra for the radio sources in the ELAIS-N1 field, and are publicly available as a part of the SDSS\,DR12.\footnote{\url{https://www.sdss.org/dr12/algorithms/ancillary/boss/sdsslofar/}}
The SDSS catalogue provides spectroscopic redshift ($z_{\rm spec}$) along with object-type classification of the spectra \citep[see][for details]{Bolton_2012}. 
The radio counterparts of the sources in the BOSS catalogue were, in part, classified based on this information.
Furthermore, the sources with reliable $z_{\rm spec}$ are identified with the flag $\texttt{ZWARNING} = 0$, and 597 sources in the uGMRT catalogue at 400\,MHz were found to have a reliable $z_{\rm spec}$. These sources were classified into AGN and SFGs, where SFGs also contain the starburst galaxies.

\subsubsection{Infrared Data}

The Spitzer Wide-Area Infrared Extragalactic (SWIRE) survey covers a sky area of 49\,deg$^2$ using the Infrared Array Camera (IRAC) at 3.6, 4.5, 5.8, 8\,$\upmu$m, and using the Multi-Band Imaging Photometer (MIPS) at 24, 70 and 160\,$\upmu$m of the \textit{Spitzer space telescope} \citep{Lonsdale_2003, Mauduit_2012}. As a part of the SWIRE survey, six extragalactic deep-fields were observed, including 8.72\,deg$^2$ on the ELAIS-N1 field \citep{Rowan_2008,Rowan-Robinson2013}. This also includes another five bands ($U', g', r', i'$ and $Z'$) from the Wide Field Survey using the 2.5-m Isaac Newton Telescope 
\citep{McMahon_2001}. The revised SWIRE catalogue includes the $J, H$ and $K_s$ bands from the Two Micron All Sky Survey (2MASS) and the UKIRT Infrared Deep SkySurvey (UKIDSS) in the near-infrared \citep{Lawrence_2007}. The availability of a large number of photometric data reduces the fraction of catastrophic outliers making the photometry in the SWIRE catalogue one of the most reliable.

In addition, we have also made use of the \textit{Herschel} Multi-tiered Extra-galactic Survey (HerMES) performed using the \textit{Herschel Space Telescope} that mapped a set of nested fields covering a total area of $\sim380$\,deg$^2$ using the \textit{Herschel}-Spectral and Photometric Imaging Receiver (SPIRE) at 250, 350 and 500\,$\upmu$m \citep{Roseboom_2010, Roseboom_2012}. SPIRE has covered 3.25\,deg$^2$ area of the ELAIS-N1 field \citep{Oliver_2012} at these bands. We have utilized these far-infrared photometry in our study.

\subsubsection{Redshifts}\label{sec_redshift}

We obtained $z_{\rm spec}$ for 597 sources (23\,per cent) in the uGMRT catalogue from the BOSS catalogue discussed earlier.
For 2216\,sources (87\, per cent) in the uGMRT catalogue, we have used the redshifts provided as a part of the LoTSS catalogue \citep{Duncan_2021}, of which 555\,sources have $z_{\rm spec}$ from BOSS. For the remaining 1661 sources, 66 sources (4\,per\,cent) have $z_{\rm spec}$ from various other spectroscopic data, and 1595 sources (95\,per\,cent) have photometric redshifts ($z_{\rm ph}$). We also found 63 additional sources in the uGMRT catalogue that have no redshift information either from BOSS or LOFAR redshift catalogue but have $z_{\rm ph}$ from the SWIRE photometric redshift catalogue \citep{Rowan-Robinson2013}. Overall, 2321 (92 per cent) of the radio sources in the uGMRT catalogue have redshift information, and these sources form the core sample of our analysis of the radio--IR relations.

\begin{figure*}
\centering
\subfloat[ Radio Luminosity \textit{versus} $z$.\label{fig:radio}]{
  \includegraphics[width=5.8cm]{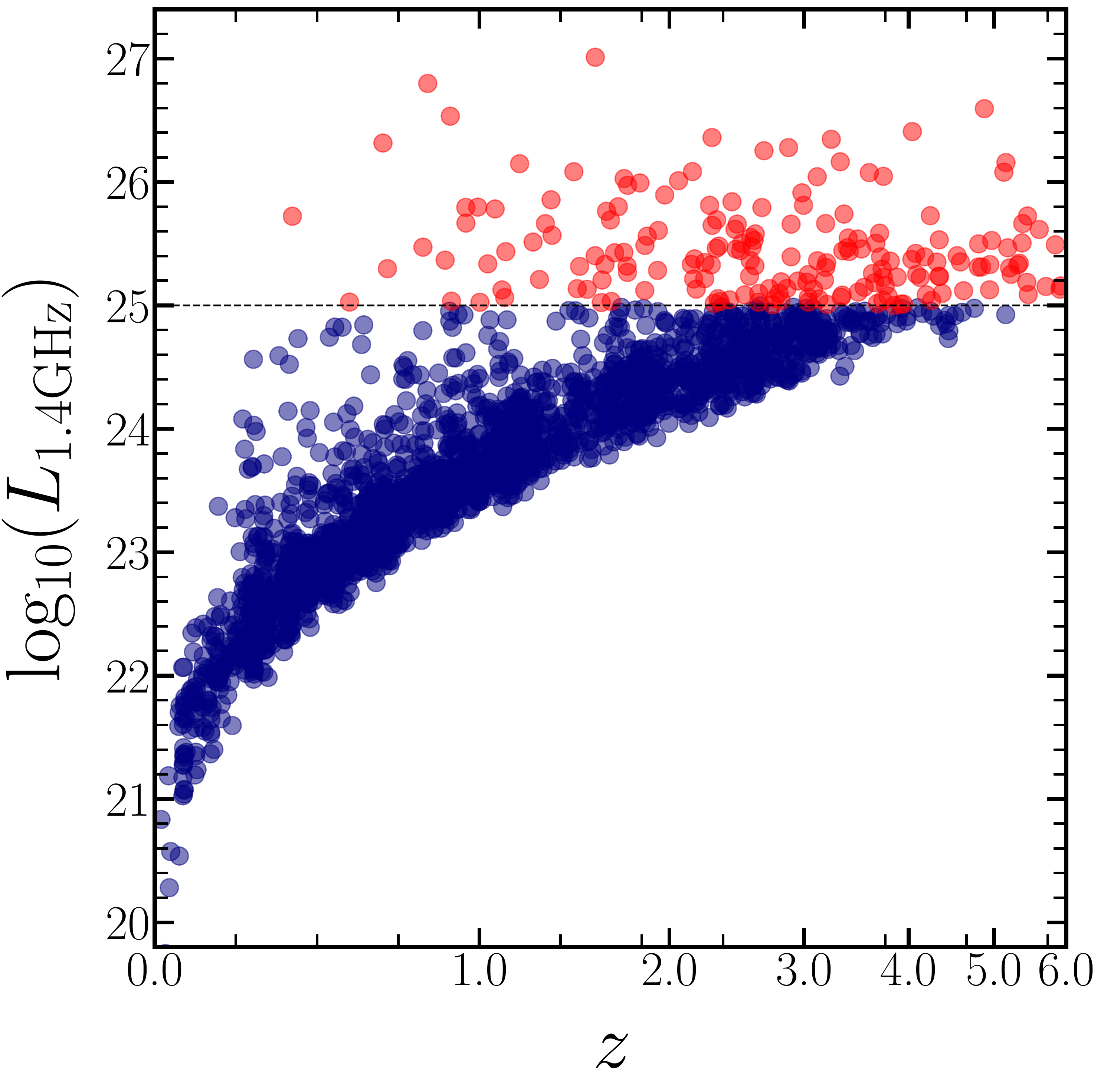}
}
\qquad
\subfloat[IRAC colour-colour distribution. \label{fig:IRAC}]{
  \includegraphics[width=5.8cm]{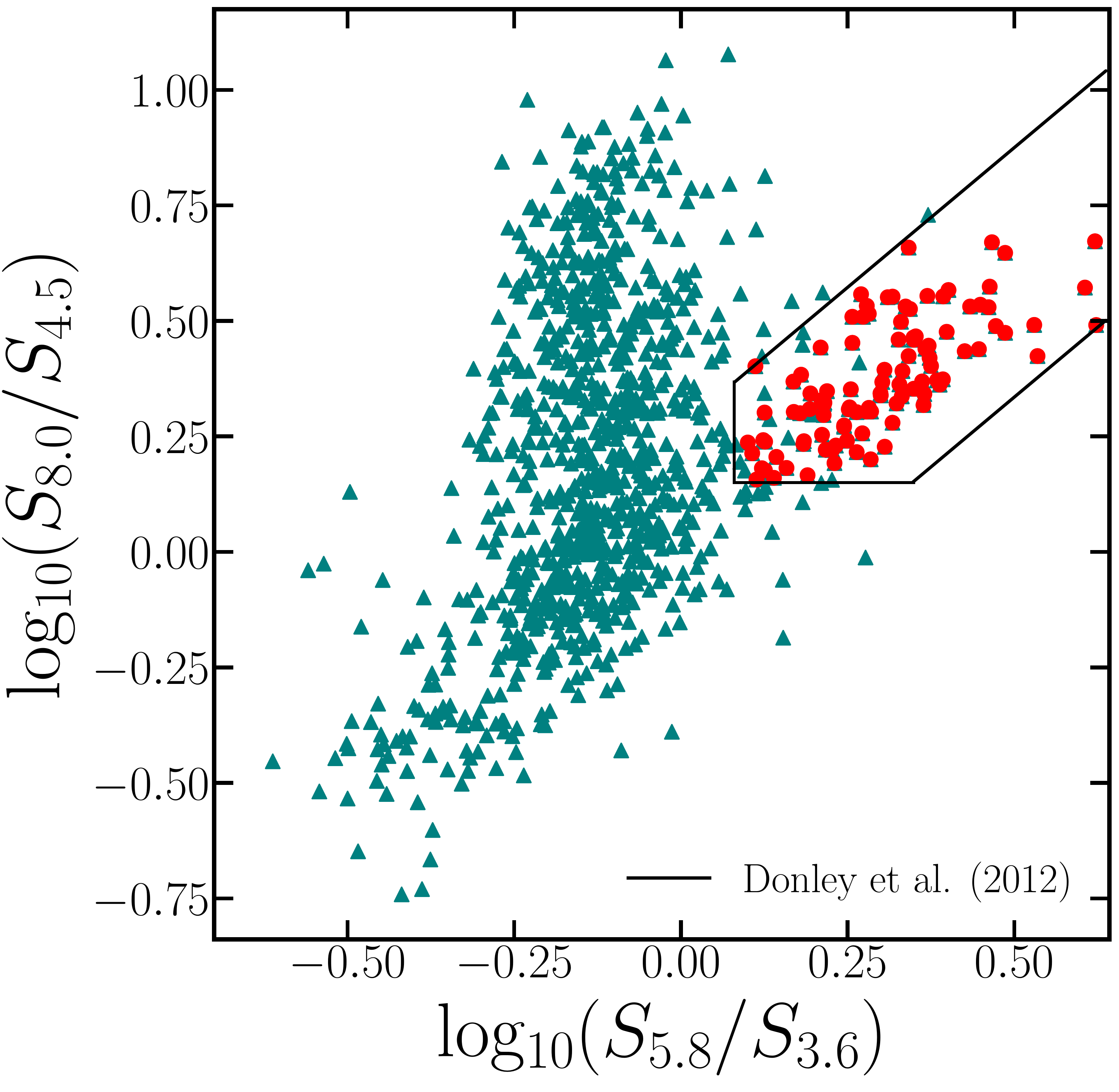}
 }
\hspace{0mm}
\subfloat[Redshift distribution for AGN and SFGs. 
\label{fig:BOSS}]{
  \includegraphics[width=5.9cm]{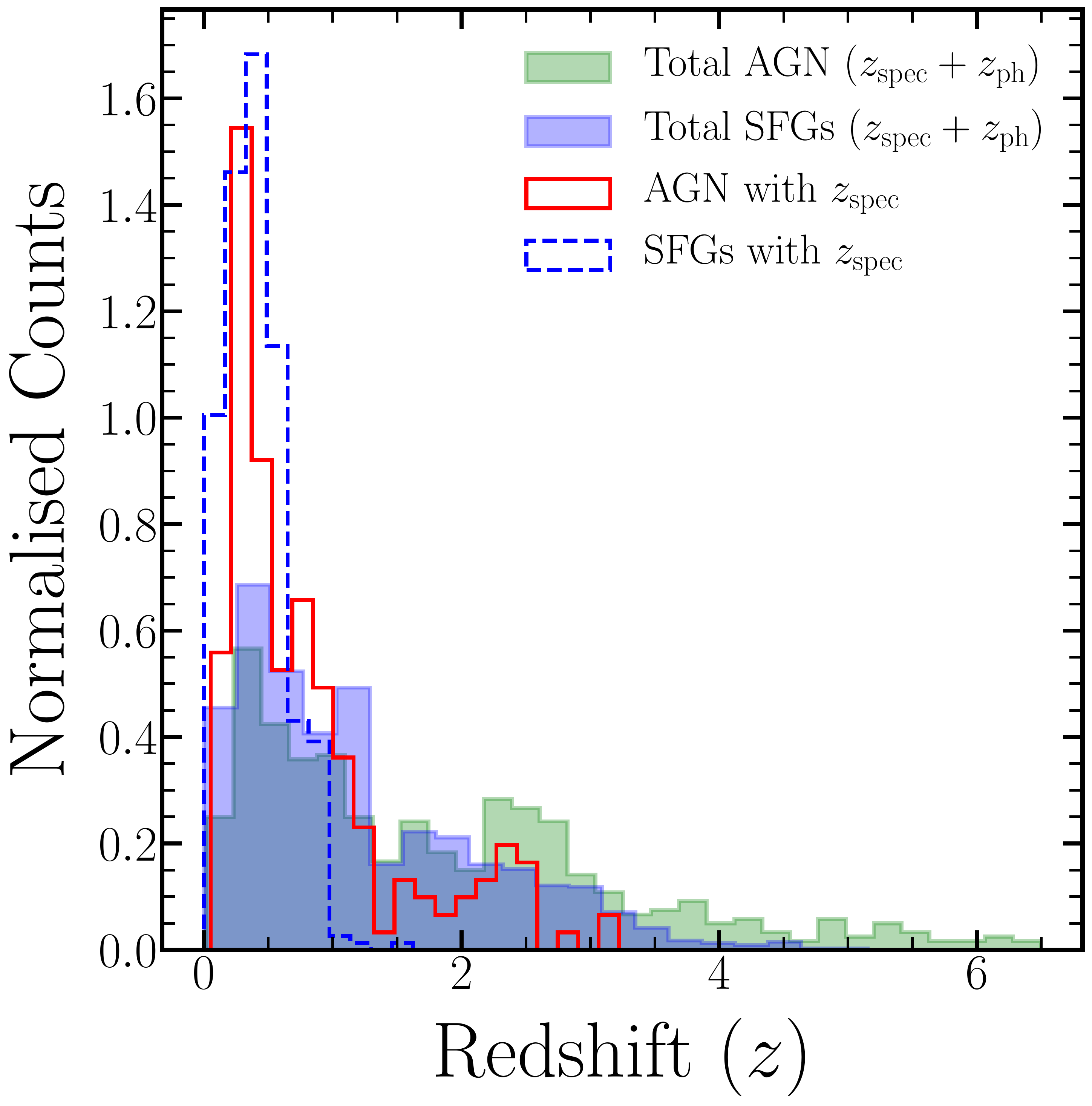}
}
\qquad
\subfloat[Plot of observed $\qmir$ values with $z$\label{fig:q24}]{
   \includegraphics[width=5.8cm]{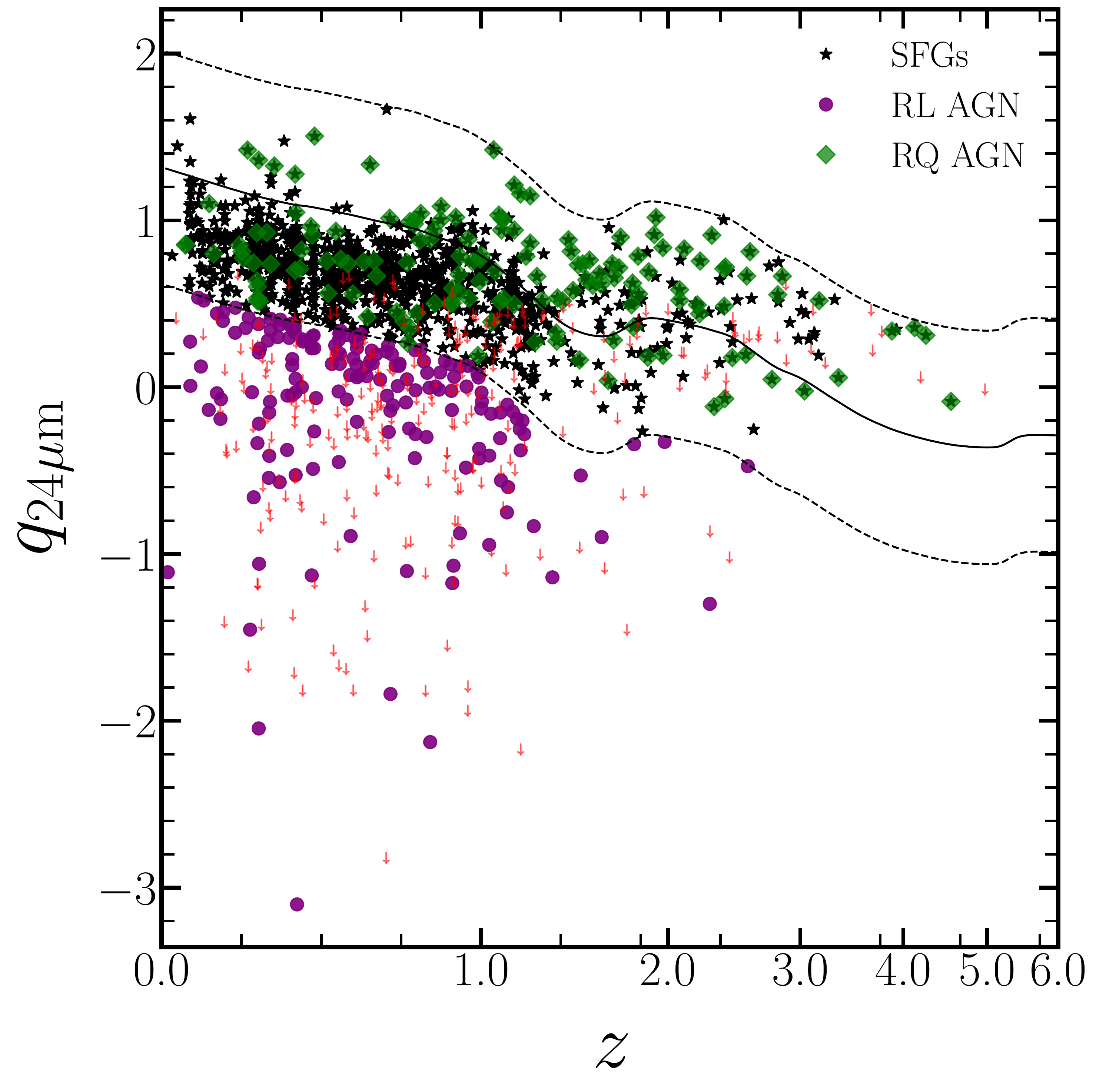}
}
\caption{Source classification schemes used in this work. (a) Variation of rest-frame luminosity at 1.4\,GHz ($\Lum$) with redshift ($z$). The dashed black line indicates the threshold $\Lum = 10^{25}\,\rm{W Hz}^{-1}$ above which the sources were classified as RL\,AGN, and are shown as the red dots (see Section~\ref{sec:radio}). (b) IRAC color-color diagram (see Section~\ref{sec:IRAC}). The black lines are from \citet{Donley_2012} and the sources within these lines represents AGN while those outside are SFGs. The red dots represent AGN, and the blue triangles represent SFGs.
(c) The redshift distribution of the sources in ELAIS-N1 field. The open histograms are spectroscopic redshift ($z_{\rm spec}$) from BOSS which provides sources classified based on their spectra. The red and the dashed blue histograms are for AGN and SFGs, respectively, identified by BOSS (see Section~\ref{sec:BOSS}). We also show the combined $z_{\rm spec}$ and photometric redshift ($z_{\rm ph}$) distribution of all the SFGs and AGN in the ELAIS-N1 identified using other classification schemes as the filled blue and green histograms, respectively. All histograms are normalized to unit area.
(d) Variation of $\qmir$ as a function of $z$ (see Section~\ref{sec:q24}). The solid black curve shows the redshifted $\qmir$ template estimated from the SED of the starburst galaxy M\,82 with it's $\pm2\sigma$ dispersion shown in black dashed curve. The black stars, green diamonds and purple dots represent SFGs, RQ\,AGN and RL\,AGN respectively. The undetected sources at 24\,$\upmu$m are shown as the upper limits and represented by the red downward pointing arrows. }
\label{fig:agn_class}
\end{figure*}

\section{AGN/SFG classification} \label{sec:agn_sfg}

Since the majority of the radio sources in the 400-MHz uGMRT data do not have spectroscopic identification, it is challenging to robustly identify the AGN and the SFGs in a radio source catalogue. The complication mostly arises due to the fact that, although continuum and/or emission lines at other wavebands, e.g., in the infrared, optical and X-ray bands can discern AGN activity, whether the AGN-component of the emission significantly dominates over the emission due to star formation activity at radio frequencies remain unclear. This is especially the case for the `radio-quiet' AGN (RQ AGN) population whose radio emission has been suggested to be dominated by star formation in the host galaxy \citep[e.g.,][]{Sopp_1991, Hodge_2008, Montenegro_2022arXiv}. To compare the properties of the radio--IR relations with SFGs, we have also studied the AGN population as well.

Following \citet{Bonzini_2013}, we have used four different source classification criteria, these are based on radio luminosity, colours in the IRAC bands, the flux density ratio at 24\,$\upmu$m in the MIR to that at 1.4\,GHz in the radio ($q$ parameter), and spectroscopy.
In addition, we have also used source classification available from the LoTSS catalogue. We denote the number of AGN and SFGs identified using these methods as $N_{\rm AGN}^\#$ and $N_{\rm SFG}^\#$ respectively, where `\#' denotes the classification scheme, `lum', `IR',`LoTSS', `$q$', and `spec', for classification based on radio luminosity, IRAC colours, LoTSS catalogue, $q$-parameter, and spectroscopy, respectively. Table~\ref{tab:classification_Schemes} summarizes the number of sources identified in both categories using these source classification schemes.

\begin{enumerate}
    \item \textit{Radio luminosity based classification: }\label{sec:radio} We computed the rest-frame luminosity at 1.4\,GHz ($\Lum$) for the 2321 sources as,
\begin{equation}
    \Lum = 4\,\uppi\,d_{\rm L}^{2} \frac{S_{1.4\,\rm{GHz}}} {(1+z)^{1+\alpha}},
    \label{eq:lum}
\end{equation}
where $S_{\rm 1.4\,GHz}$ is the observed flux density at 1.4\,GHz, $d_{\rm L}$ is the luminosity distance, and $\alpha$ is the spectral index (defined as $\boldsymbol{S\propto \nu^\alpha}$) of a given source  (see Section~\ref{sec:spec} for details). The sources with $\Lum > 10^{25}$ WHz$^{-1}$ are classified as `radio-loud' AGN (RL AGN) \citep{Sajina_2007, Sajina_2008, Jiang_2007}, as such luminosities from star formation related synchrotron emission are highly unlikely in a large population of galaxies. In Fig.~\ref{fig:radio}, we show the variation of $\Lum$ as a function of $z$, and we found 190 sources ($N_{\rm AGN}^{\rm lum}$), shown as the red points, meeting the above criterion that were classified as RL\,AGN.

    \item \textit{IRAC colour based classification:}  \label{sec:IRAC} The thermal radiation from dust, predominantly heated by the AGN, emits with a characteristic power-law spectrum at MIR wavelengths. We have used the criterion described by \citet{Donley_2012} in the IRAC colour-colour plot for 1470 sources with redshift information shown in Fig.~\ref{fig:IRAC} to identify AGN using a black wedge.
    Using the IRAC colours, we found a total of 103 AGN ($N_{\rm AGN}^{\rm IR}$) in the 400-MHz uGMRT catalogue, all of which have redshifts.

\begin{table*}
\centering
\begin{tabular}{ccccccc}
\hline 
\# & Classification & Source-type & Selection & Number of & Number of & Radio\\
& scheme & identified & criterion & AGN ($N_{\rm AGN}^\#$) & SFGs ($N_{\rm SFG}^\#$) & counterparts\\
\hline \hline \\
lum& Radio Luminosity& AGN & $\Lum \geq 10^{25}\rm WHz^{-1}$& 190 & 2129$^\dagger$ & 2319\\  
IR& IRAC Color-Color & AGN & Donley wedge & 103 & 1367$^\dagger$ & 1470\\
spec & BOSS Spectra & AGN/SFG & Spectroscopy & 82 & 513 & 597\\
LoTSS & LOFAR catalogue & AGN & opt/IR/Xray & 219 & 1997$^\dagger$  & 2216\\
q& $\qmir$ parameter & AGN & M\,82 SFG locus & 312 & 889$^\dagger$ & 1201\\
&&&&&&\\
Total unique   &    &  &     & 556 & 1763 & 2319 \\
\hline
\end{tabular}
\caption{Summary of the number of sources identified using the four classification schemes used in this work (see Section~\ref{sec:agn_class}). We find the following matches of AGN identified from radio luminosity with other classification schemes: $N_{\rm AGN}^{\rm lum} \cap N_{\rm AGN}^{\rm IR} = 8$, $N_{\rm AGN}^{\rm lum} \cap N_{\rm AGN}^{\rm q} = 14$, $N_{\rm AGN}^{\rm lum} \cap N_{\rm AGN}^{\rm spec} = 7$, $N_{\rm AGN}^{\rm lum} \cap N_{\rm AGN}^{\rm LoTSS} = 22$. Also note that we have dropped 2 sources that were classified as stars in the BOSS catalogue. Hence, we are left with 2319 sources with redshift measurements.\\ 
$^\dagger$ These represents the number of non-AGN sources from the respective criterion, and does not identify SFGs adequately. In our study, all those sources are considered as SFGs that remained identified as non-AGN after combining all the selection criteria (see text for detail).}
\label{tab:classification_Schemes}
\end{table*}

\begin{table*}
\centering
\begin{tabular}{c c c c c }
\hline \hline \\
Class &  & Number & Percentage 
& Criterion \\ [1ex]
\hline
AGN \ \ &  & 556   &   23.9 & $N_{\rm AGN}^{\rm Total} = N_{\rm AGN}^{\rm lum} \cup N_{\rm AGN}^{\rm IR} \cup N_{\rm AGN}^{\rm q} \cup N_{\rm AGN}^{\rm LoTSS} \cup N_{\rm AGN}^{\rm spec}$\\
 & RQ AGN & 146 &  6.3 & $N_{\rm AGN}^{\rm RQ} = (N_{\rm AGN}^{\rm lum} \cap N_{\rm SFG}^{\rm q}) \cup (N_{\rm AGN}^{\rm IR} \cap N_{\rm SFG}^{\rm q}) \cup (N_{\rm AGN}^{\rm spec} \cap N_{\rm SFG}^{\rm q}) \cup (N_{\rm AGN}^{\rm LoTSS} \cap N_{\rm SFG}^{\rm q})$  \\
 & RL AGN & 333 &  14.3 & $N_{\rm AGN}^{\rm RL} = (N_{\rm AGN}^{\rm lum} - (N_{\rm AGN}^{\rm lum} \cap N_{\rm SFG}^{\rm q})) \cup (N_{\rm AGN}^{\rm q} - N_{\rm AGN}^{\rm RQ})$ \\ \\
 \hline \\
SFGs \ \ &  & 1763  &   76.0 & $N_{\rm SFG}^{\rm Total} = 2319 - N_{\rm AGN}^{\rm Total}$ \\
\hline
\end{tabular}
\caption{Total unique number of SFGs and AGN identified from the selection criterion discussed in Section~\ref{sec:agn_class}. We point out that we have not made any distinction between RL and RQ\,AGN, and commonly refer to them as AGN in this paper.}
\label{tab:classification}
\end{table*}

    \item \textit{Spectroscopic classification:} \label{sec:BOSS} As discussed in Section~\ref{sec_redshift}, the 597-sources with $z_{\rm spec}$ from the BOSS catalogue also contain classified sources based on their spectra. We used the \texttt{CLASS} and \texttt{SUBCLASS} keywords in the BOSS catalogue for identifying AGN and SFGs which include starburst galaxies \citep[see][for details]{Bolton_2012}. From BOSS spectroscopy, we found 533 sources that were classified as GALAXY and 62 as QSOs. The remaining two sources were classified as STAR and were not included in our further analysis. Of the 533 sources with \texttt{CLASS} GALAXY, 119 were sub-classified as STARFORMING (includes one STARFORMING BROADLINE source), 102 as STARBURST, and 20 as AGN (includes \texttt{SUBCLASS}es BROADLINE and AGN BROADLINE). The remaining 292 sources had no sub-classification. In summary, we identify a total of 82 AGN ($N_{\rm AGN}^{\rm spec}$), and 513 SFGs ($N_{\rm SFG}^{\rm spec}$) from BOSS spectroscopy (see Table~\ref{tab:classification_Schemes}).

    \item \textit{LOFAR based classification:} The LOFAR photometric redshift catalogue \citep{Duncan_2021} also contains sources classified based on various multi-wavelength information found in the literature. This catalogue includes AGN identified from the  Half Million Quasars (HMQ) catalogue \citep{Flesch_2015},  IRAC color--color using \citet{Donley_2012}, and X-ray catalogue using \citet{Boller_2016}. We used the flag \texttt{AGN} in the LOFAR redshift catalogue for selecting AGN. In this way, we identify 219 AGN ($N_{\rm AGN}^{\rm LoTSS}$) in our uGMRT sample.

    \item \textit{$\qmir$ based classification:} \label{sec:q24} In addition to the above classification schemes, for the 1201 sources detected at $24\,\upmu$m, we have also used the logarithmic ratio of the observed mid-IR at $24\,\upmu$m and radio flux densities, $q_{24\, \upmu \mathrm{m}} = \mathrm{log}_{10}(S_{24\, \upmu \mathrm{m}}/ S_{\rm1.4\,GHz})$, to identify AGN. Here, $S\rm_{24\upmu m}$ is the observed flux density at 24$\upmu$m. In Fig.~\ref{fig:q24} we show the plot of observed $\qmir$ \textit{versus} redshift for the uGMRT-selected sources. We find a median $\qmir = 0.62\pm0.20$ using the observer's frame flux densities with median $z = 0.61$. Our value of median $\qmir$ in the observer's frame obtained by extrapolating the 400\,MHz flux densities to 1.4\,GHz matches excellently with those reported in the literature by using observations at 1.4\,GHz \citep[e.g.,][]{Appleton_2004, Huynh_2010, Ibar_2008}. We have utilized the redshifted $\qmir$ values from the SED template of the nearby starburst galaxy M\,82 to differentiate  SFGs and RL AGN as described in \cite{Bonzini_2013}. The M\,82 template was normalized to the local average value of $q\rm_{24\, \upmu m}$ by \cite{Sargent_2010_2} which is shown as the solid black curve in Fig. \ref{fig:q24}. The SFG locus is defined as the region within $\pm 2\,\sigma$ of the M\,82 template shown by the black dashed curves in Fig.~\ref{fig:q24}, where $\sigma$ is the typical scatter of 0.35\,dex.

In summary, we classified a source as RL AGN if it has $\Lum > 10^{25}\,\whz$ or if the $\qmir$ value lies $2\,\sigma$ below the M\,82 template curve. A source is classified as a radio-quiet AGN (RQ\,AGN) if an AGN identified using any of the other criteria lies within the SFG locus. Besides the SFGs that have spectroscopic identification in the BOSS catalogue, we consider the remaining sources within the SFG locus as SFGs for this work. Using the observer's frame $\qmir$ values, we identify 166 sources as RL\,AGN, 146 sources as RQ AGN, and the remaining 889 as SFGs, as described above. 
Thus, based on observed $\qmir$ values, we identify 312 sources ($N_{\rm AGN}^{\rm q}$) as AGN that comprise both RQ and RL AGN obtained using this criterion. The black stars, purple dots, and green diamonds in Fig.~\ref{fig:q24} represent SFGs, RL\,AGN, and RQ\,AGN, respectively. We emphasize that we have not made any distinction between RL and RQ\,AGN in the rest of the paper, and both of them are referred to as AGN. 
For the remaining 283 sources with redshift measurements from SWIRE catalogue but having no counterpart at $24\,\upmu$m, we have used the upper limits on their $\qmir$, and are shown as the downward pointing arrows in Fig.\,\ref{fig:q24}.
It is clear that a bulk of these sources lie in the RL\,AGN regime ($\sim 80$\,per\,cent of the undetected sources at $24\,\upmu$m) as per the redshifted SED template of M\,82.

In Table~\ref{tab:classification}, we summarize the number and fraction of AGN and SFGs with respect to the total of 2319 sources that were classified. The redshift distributions of AGN and SFGs identified from different methods are shown in Fig.~\ref{fig:BOSS}. The distribution of $z_{\rm spec}$ from BOSS for AGN and SFGs are shown by the open histograms, and the distributions for $z_{\rm spec}$ and $z_{\rm ph}$ combined are shown by the shaded histograms, where the median redshift of AGN and SFGs in our sample are found to be 1.45 and 0.94, respectively. 

\end{enumerate}

\begin{figure*}
    \centering
    \includegraphics[width= 7cm]{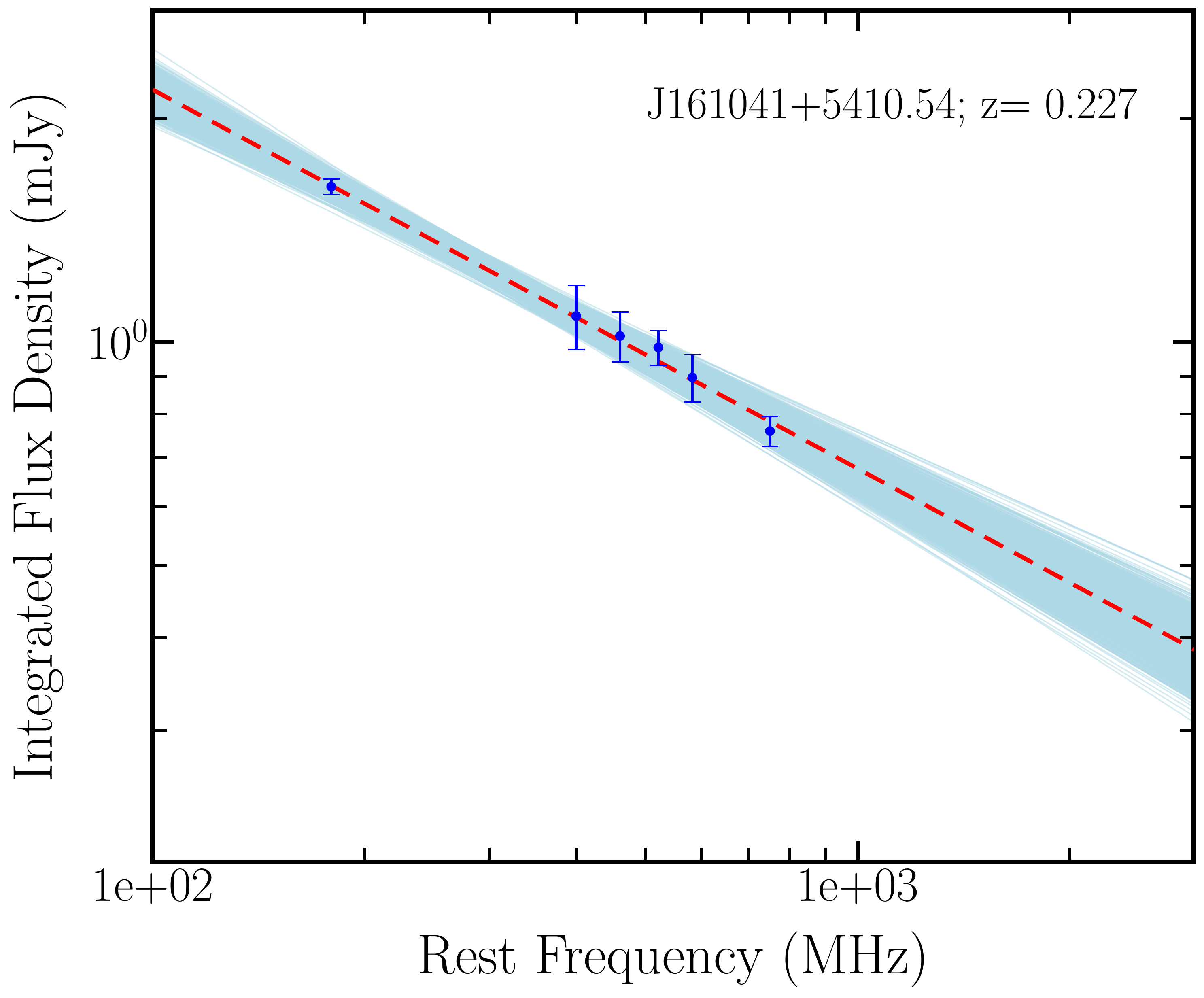}
    \includegraphics[width= 7cm]{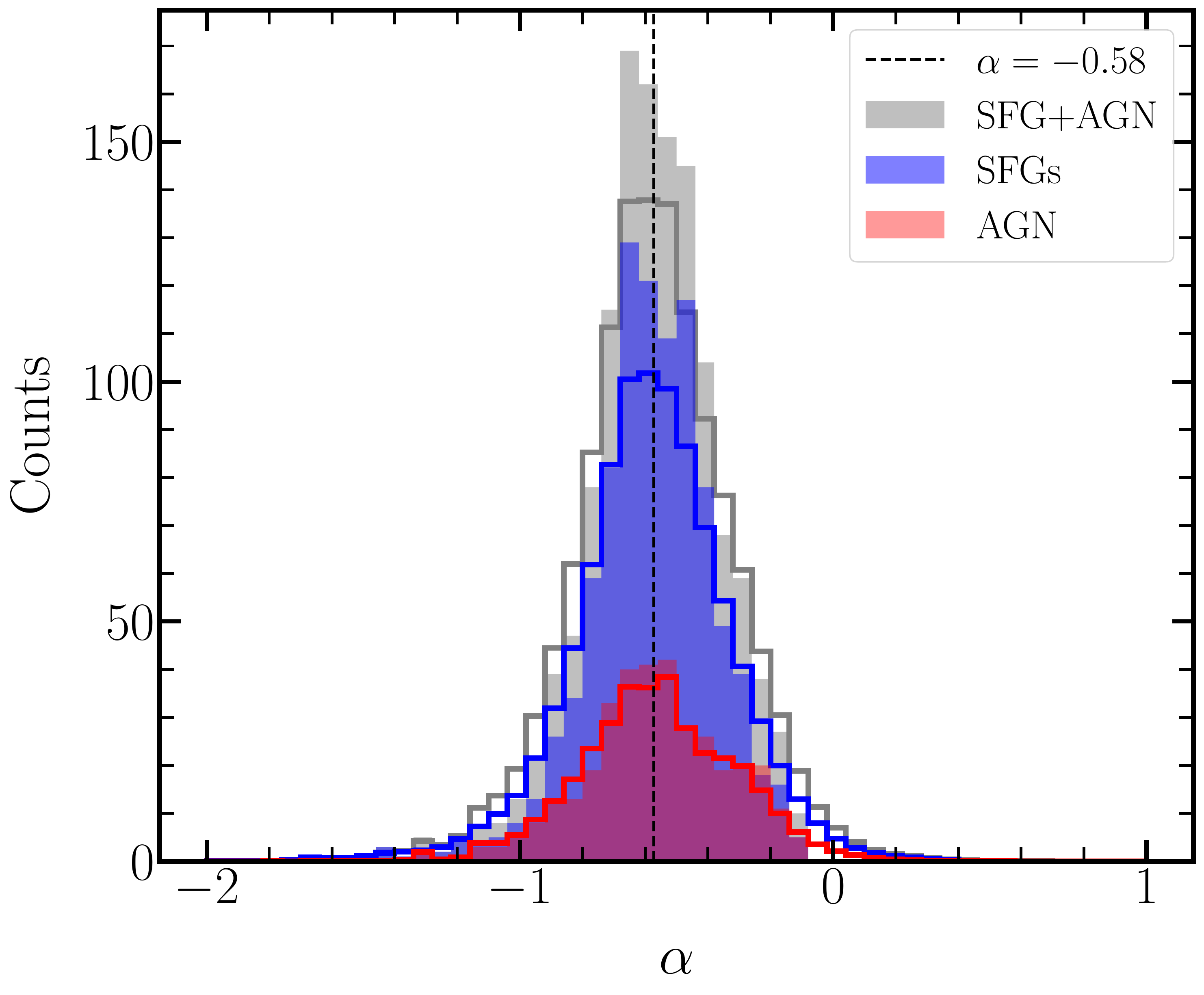}
    \caption{\textit{Left}: Typical rest-frame radio SED fitted for one of the sources J161041+5410.54 at $z = 0.227$. The errorbars are the $3\sigma$ errors on the flux densities. The light blue lines show the power-law fit for 1000 samples randomly drawn with the $3\sigma$ flux density error at each frequency (see section~\ref{sec:spec} for details). The red dashed line represents the best-fit SED for this source.
    \textit{Right}: The spectral index ($\alpha$) distribution of the radio sources in the ELAIS-N1 field obtained by fitting the power-law shown on left for 1278 sources (see section~\ref{sec:spec}). The shaded histograms represent the distribution for SFGs, AGN and for the combined population (SFG+AGN). The dashed line represents the median alpha of $-0.58$ for all the sources. The open histograms show the distribution by accounting for the error in the estimated $\alpha$ for each source. The median $\alpha$ then measured for SFGs and AGN are $-0.58$ and $-0.57$, respectively.}
    \label{fig:SED and alpha_dist_new}
\end{figure*}

\subsection{Salient features of the classification schemes}\label{sec:agn_class}

In Table~\ref{tab:classification_Schemes}, we summarize the various methods we have used for classifying the radio sources in the ELAIS-N1 field into AGN and SFGs. Note that the number of SFGs ($N_{\rm SFG}^{\#}$) represent those sources that were not identified as AGN from the different criteria discussed in the previous section, except for the ones that were identified as SFGs from BOSS spectroscopy. We would like to emphasize that the statistical selection of SFGs in this way could possibly mis-classify AGN in composite systems as SFGs or \textit{vice versa}. Larger gas content in star-forming low-excitation radio galaxies (LERGs) marks an increase in the population at high redshifts \citep{Williams_2018}. Hence, the population of SFGs in our sample could be contaminated by LERGs at $z\gtrsim0.7$. As described in the previous section, we identify 146 sources as RQ AGN that indicate emission from AGN identified from the other four selection criteria (lum, IR, LoTSS, and spec), and lie within the M\,82 SFG locus. Further, a source is considered as RL\,AGN if it was classified as AGN either from radio luminosity or from the observer's frame $\qmir$ values. In this way, we classify an overall of 333 sources as RL\,AGN. There are 77 AGN that were neither classified as RQ or RL\,AGN. Combining all the five source classification criteria, we find a total of 580 ($\sim25$\,per cent) AGN (146 RQ\,AGN + 333 RL\,AGN + 77 AGN), while the rest 1763 (76\,per cent) sources that have redshift measurements but do not show AGN characteristics are considered as SFGs in our study.

Interestingly, only 10 out of 190 RL\,AGN classified based on $\Lum$ has been detected in BOSS, of which 7 are identified as AGN from BOSS spectroscopic classification. Out of these, 4 sources were identified as AGN from IRAC colors. Additionally, 8 sources were identified as AGN using the $\Lum$ criteria and the IRAC colours. On the other hand, 14 of the 190 RL\,AGN identified using radio luminosity are also identified as RL\,AGN from the $\qmir$ criteria. In general, from Fig.~\ref{fig:radio} we notice that RL\,AGN are largely missing in our radio observations, especially at $z\lesssim 0.2$. This is partly due to the relatively small co-moving volume probed by our uGMRT observations and flattening of the radio luminosity function of AGN at low redshifts \citep{Mauch2007}. Furthermore, the AGN population is dominated by LERGs at low redshifts \citep{Heckman_Beck_2014, Hardcastle&Croston_2020} making their detection challenging in large sky-area spectroscopic surveys, such as the BOSS. In contrast, spectroscopy tends to identify high-excitation radio galaxies (HERGs) via their strong high-excitation emission lines at high redshifts. Furthermore, from Fig.~\ref{fig:q24} it is clear that several sources that are undetected at $24\,\upmu$m, falls under the RL\,AGN category. A combination of relative faintness and dust obscuration at near- and mid-infrared wavebands perhaps makes them undetectable in the SWIRE catalogue and explains the relatively low number of RL\,AGN that are identified by the classification schemes we have used. These RL\,AGN are hence missed when we study the radio--IR relations in later sections. We again emphasize that no distinction is made between RL and RQ\,AGN henceforth, and both types are considered as AGN in our study.

\section{Correction to rest-frame} \label{sec:radio_IR_SED}

In this section, we present the $k$-correction methods we have applied to the observed flux densities at the radio and infrared wavebands for determining the respective rest-frame emissions. For this, we use only those sources that have redshift information.

\subsection{\textit{k}-correction at radio frequencies}\label{sec:spec}

To investigate the radio--IR relations, we determined the luminosity of the radio sources at 1.4\,GHz and 400\,MHz in the rest-frame. For this, we constructed the radio SED by taking advantage of the 200\,MHz wide bandwidth of the uGMRT data, and by using flux densities measured at 146\,MHz, 612\,MHz and near 1.4\,GHz, wherever available. To obtain a reliable spectral index ($\alpha$), we model the radio SED between the frequency range 100\,MHz and 1.4\,GHz as a power-law of the form, 
\begin{equation}
S(\nu) = S_{\rm 1.4\,GHz}\,\left(\frac{\nu}{1.4\,{\rm GHz}}\right)^\alpha, 
\label{eq:radioSED}
\end{equation}
where $\nu$ is expressed in GHz.

To achieve a robust radio SED modeling, we first divided the uGMRT data, covering the 300--500\,MHz frequency range, into narrower 50\,MHz bandwidth centered at 325, 375, 425, and 475\,MHz. Each of these narrow bandwidth data were imaged using the task \texttt{tclean} available as a part of the \texttt{CASA} package.\footnote{\url{http://casa.nrao.edu/} \citep{McMullin2007}.} Using P{\tiny Y}BDSF on each image individually, we compiled four catalogues at 325, 375, 425, and 475\,MHz, wherein we obtained 1513, 1584, 2199, and 1366 sources, respectively, above $5\sigma$.

Out of the 2321 sources in the 400-MHz uGMRT catalogue with redshift information, there are 1278 sources for which we could perform SED fitting by ensuring at least 3 data points at any of the 7 frequencies be available for a source.
In order to robustly account for the flux density errors while fitting, we performed Monte-Carlo simulations with 1000 random realizations of the flux densities at each frequency, drawn within $3\sigma$ error, for a source. Each realization was fitted using equation~\ref{eq:radioSED}. For each source, we estimated the value of $S_{\rm 1.4\,GHz}$ and $\alpha$, and their respective errors, as the corresponding mean, and standard deviation from the 1000 realizations. In the left-hand panel of Fig.~\ref{fig:SED and alpha_dist_new}, we show the SED in the rest-frame for one of the sources, J161041+5410.54 at $z = 0.227$. The power-law fit for each of the 1000 samples are shown as the light blue lines. The red dashed line represents the best-fit SED. 
The median $\alpha$ for these 1278 sources was found to be $-0.57 \pm 0.01$ and a median absolute deviation (MAD) of 0.12.

In summary, out of the 2321 sources in the 400\,MHz uGMRT catalogue that have redshift information, we have obtained $\alpha$ for 1278 sources, measured using power-law SED fitting.
In the right-hand panel of Fig.~\ref{fig:SED and alpha_dist_new}, we show the distribution of $\alpha$ of these 1278 sources as the grey shaded histogram. The distributions of 351 AGN and 927 SFGs are shown as the red and blue shaded histograms. For the overall sample of 1278 sources, we find that the median $\alpha = -0.57 \pm 0.005$ having a dispersion of 0.12. For the SFGs, we find the median $\alpha = -0.57 \pm 0.006$ with a dispersion of 0.12, and for AGN, the median $\alpha = -0.57 \pm 0.008$ that has a dispersion of 0.14. In order to accurately account for the error on individual estimates of $\alpha$ for determining its statistical properties, we have drawn 1000 random samples of $\alpha$ within their respective error for each of the 1278 sources. The distribution of $\alpha$ for these 1000 random samples is shown as the open histograms in Fig.~\ref{fig:SED and alpha_dist_new} (right-panel). From this, for all the sources, we find the median $\alpha = -0.58 \pm 0.15$, wherein for SFGs the median $\alpha = -0.58\pm0.15$, and for AGN the median $\alpha = -0.57\pm0.16$. Here, the errors on the median values are the MAD measured from the distribution.
For the remaining 1043 sources that have redshift measurements but do not have enough data points to perform SED fitting, we have assumed a constant spectral index of $\alpha=-0.7$ to estimate $\Lum$ for classification in Section~\ref{sec:agn_sfg} only.
We would like to point out that for our further analyses, we have used only those sources that have $\alpha$ measured from the SED fits to keep the statistical population the same throughout the paper.

\begin{figure}
    \centering
    \includegraphics[width =7cm]{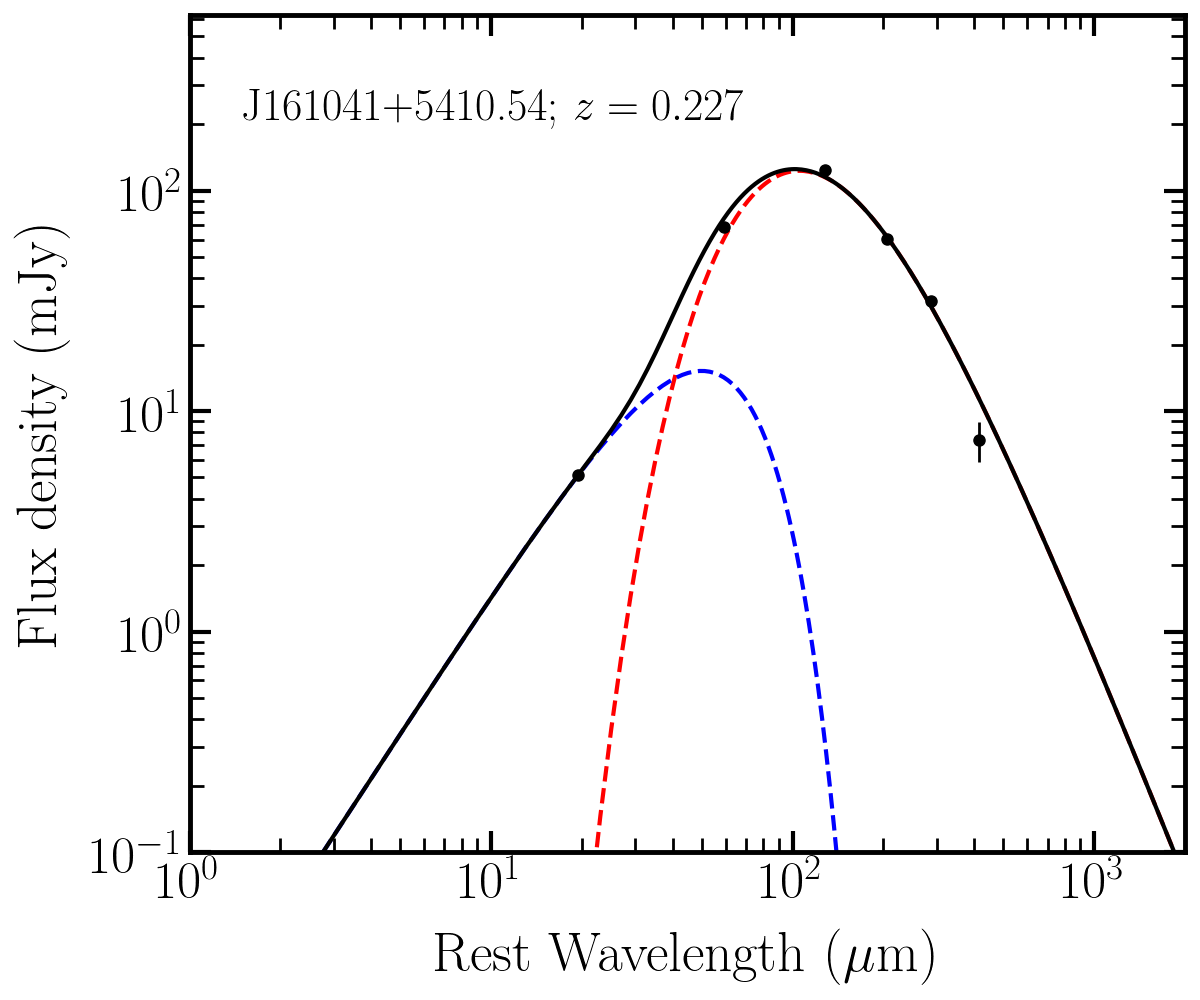}
    \caption{Typical rest-frame infrared SED of one of the source, J161041+5410.54 at $z = 0.227$ in our sample. We have modelled the infrared SED as a combination of modified blackbody shown as the red dashed curve and a mid-infrared power-law with a cut-off shown as the blue-dashed curve (see Section~\ref{sec:IR_kcorr}). The black solid line shows the combination of these components that provides an excellent fit to the data. }
    \label{fig:IR_SED}
\end{figure}

\begin{figure*}
    \centering
    \begin{tabular}{cc}
    \includegraphics[scale = 0.25]{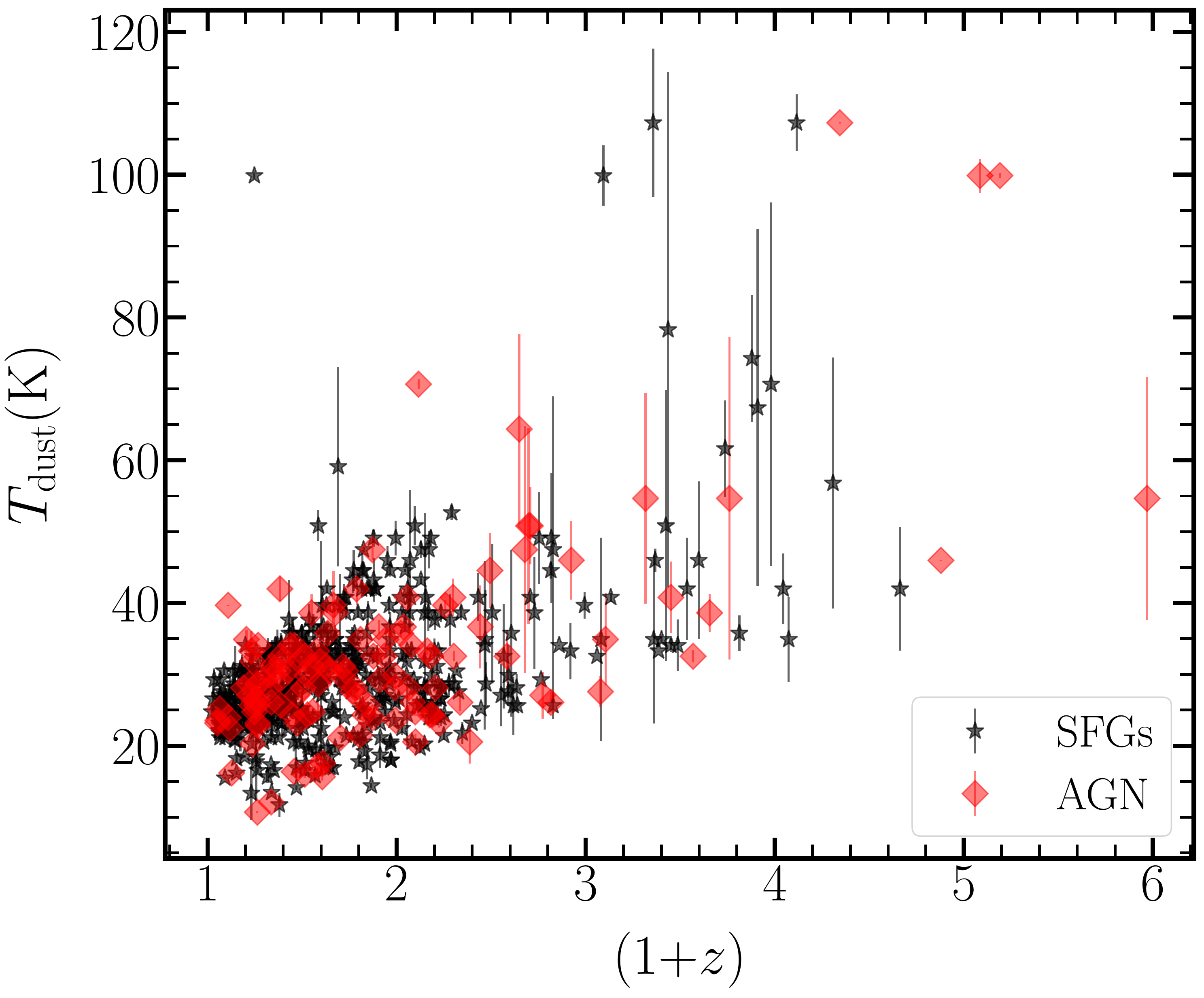}&
    \includegraphics[scale = 0.25]{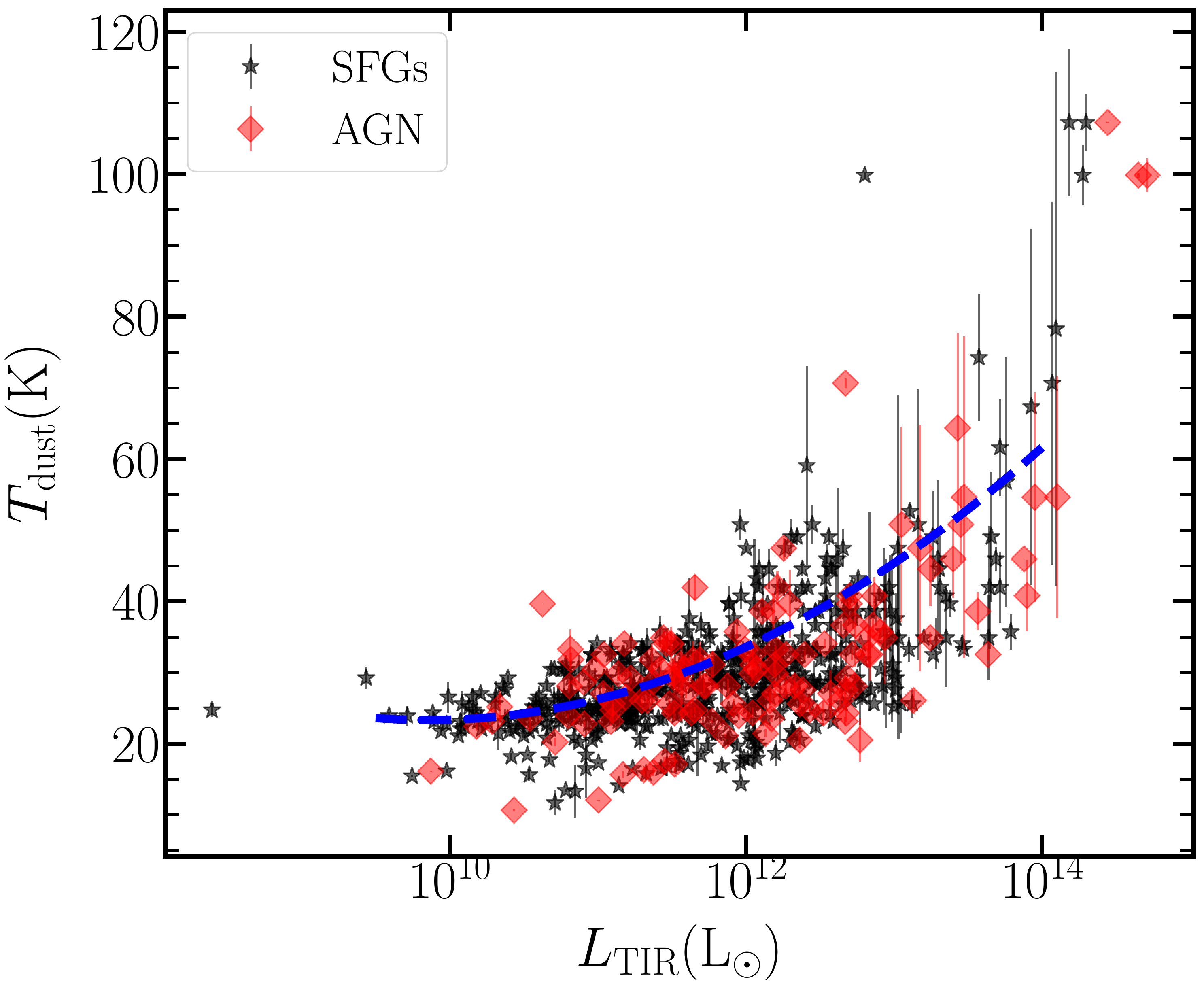}
    
    \end{tabular}
    \caption{ \textit{Left}: Variation of dust temperature ($\tdust$) with redshift $(1+z)$.
    \textit{Right}: Variation of $\tdust$ with the total infrared luminosity ($\ltir$) integrated between 8 and $1000\,\upmu$m. The blue dashed line shows the second order polynomial from \citet{Magnelli_2014}. The SFGs and AGN are shown as black stars and red diamonds, respectively, in both the panels.}
    \label{fig:Tdust}
\end{figure*}

\subsection{\textit{k}-correction at infrared wavelengths}\label{sec:IR_kcorr}

In this study we examine the radio--IR correlation between the rest-frame IR emission (both monochromatic and bolometric) and the radio emission. 
We match our radio catalogue with the SWIRE and the HerMES catalogues to obtain a total of 634 sources for which we fit the infrared SED as described below. 

For estimating the monochromatic and bolometric luminosites at infrared waveband, we model the infrared SED using a composition of single temperature modified-blackbody (graybody) and a truncated mid-IR power-law \citep{Casey_2012}. We chose this composite model of the infrared SED because a single temperature graybody spectrum does not fit the mid-IR observations well, while multi-temperature graybody SED model introduces several free parameters. The infrared SED is modeled as \citep{Casey_2012},
\begin{equation}
    S(\lambda) = A_{\rm GB}\,\frac{\left(1-{\rm e}^{-\tau_\lambda}\right)\,\lambda^{-3}}{\left({\rm e}^{h\,c/\lambda\,k\,T} -1 \right)} + A_{\rm PL}\,\left(\frac{\lambda}{\lambda_{\rm c}}\right)^{\alpha_{\rm IR}}\, {\rm e}^{-\left(\lambda/\lambda_{\rm c}\right)^2}.
    \label{eq:SED_IR}
\end{equation}
Here, $A_{\rm GB}$ and $A_{\rm PL}$ are the amplitude normalization for graybody and mid-IR powerlaw, respectively; $\lambda_{\rm c}$ is the mid-IR turnover wavelength; $\alpha_{\rm IR}$ is the mid-IR powerlaw index; $\tau_\lambda = (\lambda_0/\lambda)^\beta$ is the dust optical depth and has a value of unity at $\lambda_0 = 200\,\upmu$m; $\beta$ is the dust emissivity index assumed to be constant with $\beta=1.5$ \citep{Casey_2012, Magnelli_2014}; $T$ is the temperature; and $h$, $c$ and $k$ are the Planck constant, speed of light, and Boltzmann constant, respectively. The parameters $A_{\rm PL}$ and $\lambda_{\rm c}$ are coupled to the rest \citep[see][]{Casey_2012}, and for the assumed value of $\beta=1.5$, the number of free parameters reduces to three, namely, $A_{\rm GB}$, $T$, and $\alpha_{\rm IR}$. Therefore, at least four data points in the infrared waveband were used to constrain the SED in the mid- to far-infrared wavelengths. Further, in order to constrain the mid-IR power-law, we ensured that all sources be detected at $24\,\upmu$m in the SWIRE catalogue. In Fig.~\ref{fig:IR_SED}, we show a typical SED in the rest-frame infrared waveband for the same source, J161041+5410.54, shown in Fig.~\ref{fig:SED and alpha_dist_new} (left-hand panel). The graybody and the mid-IR power-law components are shown as the red dashed and the blue dot-dashed curves, respectively, and the solid black curve shows the total SED model. Note that, the effective dust temperature ($T_{\rm dust}$) was obtained by using the Wien's displacement law, $\tdust = b/\lambda_{\rm peak}$, where $b = 2.898 \times 10^3\,\upmu$m\,K, and $\lambda_{\rm peak}$ (in $\upmu$m) is the wavelength of the peak of the SED.

In order to determine the $k$-corrected bolometric luminosity for a source, we integrated the fitted SED between the rest-frame wavelength range $8$ and $1000\,\upmu$m to obtain the total infrared (TIR) flux, and converted them to luminosity $L_{\rm TIR}$. On the other hand, monochromatic flux densities were obtained at $24$ and $70\,\upmu$m from the best-fit SED in the rest-frame, and were converted to $L_{\rm 24\,\upmu m}$ and $L_{\rm 70\,\upmu m}$, respectively.

In the left-hand panel of Fig.~\ref{fig:Tdust}, we show the variation of $\tdust$ with $(1+z)$, and find $\tdust$ to increase with redshift for both SFGs and AGN in our sample. A linear increase of $T_{\rm dust}$ with redshift was reported by \citet{Kovacs_2006}
for sub-millimetre galaxies in the redshift range $1.5\textrm{--}3.5$, and by \citet{Basu_2015} for `blue cloud' galaxies up to $z=1.2$ in the \textit{XMM}-LSS field. This increasing trend could be caused due to $T_{\rm dust}$ being correlated with the total infrared luminosity, as shown in the right-hand panel of Fig.~\ref{fig:Tdust}, and which, in turn, is correlated with redshift due to flux-limited surveys. Overall, there are 634\,sources (520 SFGs and 114 AGN) for which we have obtained the rest-frame infrared luminosities. For 450 of these sources (349 SFGs and 101 AGN), we have measured values of $\alpha$ using multiple radio frequencies, and these sources form the core sample in our study of the radio--IR relations in the next sections. We would like to emphasize that, although the parent sample is drawn from the 400-MHz uGMRT data, these 450 sources used for further analysis have a complicated selection function due to the requirement imposed for the radio and the infrared SED fitting.

\section{Results}\label{sec:result}

Here, we focus on the properties of the radio--IR relations using the rest-frame emission for the SFGs and AGN detected in the deep uGMRT observations of the ELAIS-N1 field at 400\,MHz. In the radio waveband, we use the monochromatic rest-frame emission at 1.4\,GHz and 400\,MHz; and, to our knowledge, for the first time we will use the bolometric radio emission in high-redshift sources. In the infrared, we have used different measures of luminosity, i.e., monochromatic luminosities in the MIR at $24\,\upmu$m ($\lmir$) and in the FIR at $70\,\upmu$m ($\lfir$); and the total infrared luminosity integrated between 8 and $1000\,\upmu$m ($\ltir$). 
We first present our results obtained for the `$q$' parameter, defined as the logarithmic ratio of the $k$-corrected luminosity at infrared ($L_{\rm IR}$) to that at radio frequencies ($L_\nu$), and is given as,
\begin{equation}
 q_{\rm IR} = \log_{10}\left(\frac{L_{\rm IR}}{L_{\rm \nu}}\right).  
\end{equation}
Here, ${\rm IR} = 24\,\upmu$m, $70\,\upmu$m and TIR, and $\nu$ is the radio frequency. In this analysis, we will present the statistical properties of the different types of $q_{\rm IR}$ to study the radio-IR relations. We then study the slope $b$ of the radio--IR relations given as $L_{\nu} \propto L_{\rm IR}^{b}$ in the log--log space.

\subsection{Variation of monochromatic `$q$' parameter} \label{sec:q}

The `$q$' parameter is often used in the literature to study the evolution of the radio--IR relations.
At MIR wavelengths, $\qmir$ is used for discerning dominant emission from AGN which gives rise to an excess in the radio emission \citep[e.g.,][]{Padovani_2011,Bonzini_2013}, and thereby, lower $\qmir$ as compared to the emission from the star-forming counterparts. Once the SFG population in a sample is identified, the radio--MIR relation can be used to calibrate the radio emission to trace star formation rates in high redshift galaxies \citep{Madau_2014,Murphy_2011}. At MIR ($\sim10$--$30\,\upmu$m), however, the emission could be contaminated by cirrus dust heated by old stars and/or an obscured AGN, or emission from polycyclic aromatic hydrocarbons (PAHs). In that case, $\qfir$ or $\qtir$ are alternatively used. For a modified graybody-type SED, the IR emission in these bands are dominated by cold $\sim20$\,K dust emission in star forming galaxies, and can also be used to calibrate the radio emission to infer star formation rates \citep{Bell_2003, Murphy_2011, Tabatabaei_2017}. In this section, we will investigate the statistical properties of $\qmir$ and $\qfir$, and their variation with redshift.

\subsubsection{$q_{\rm 24\upmu m}$} \label{sec:q24variation}

\begin{figure*}
    \centering
   \includegraphics[height=7cm]{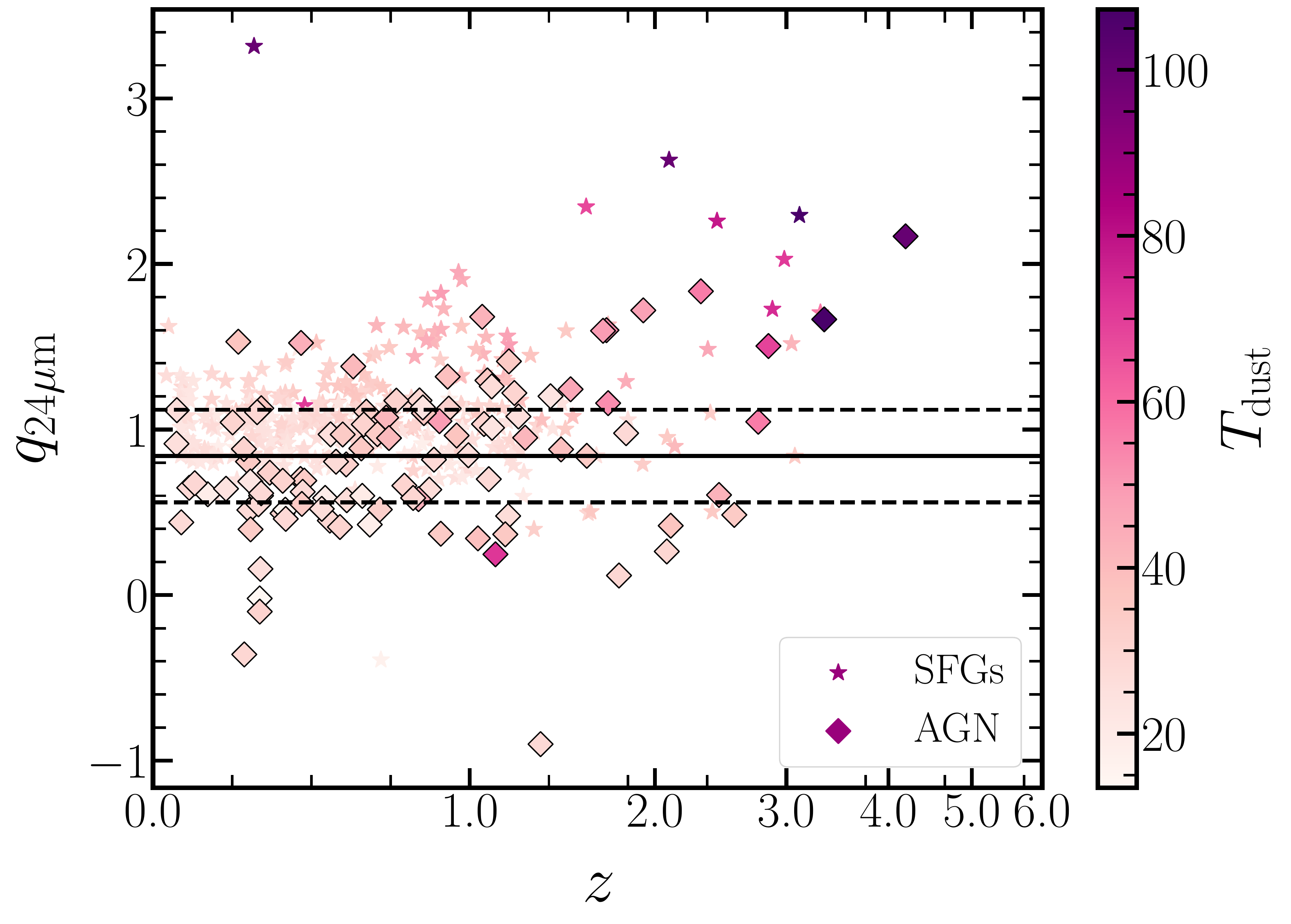}
   \includegraphics[height=7cm]{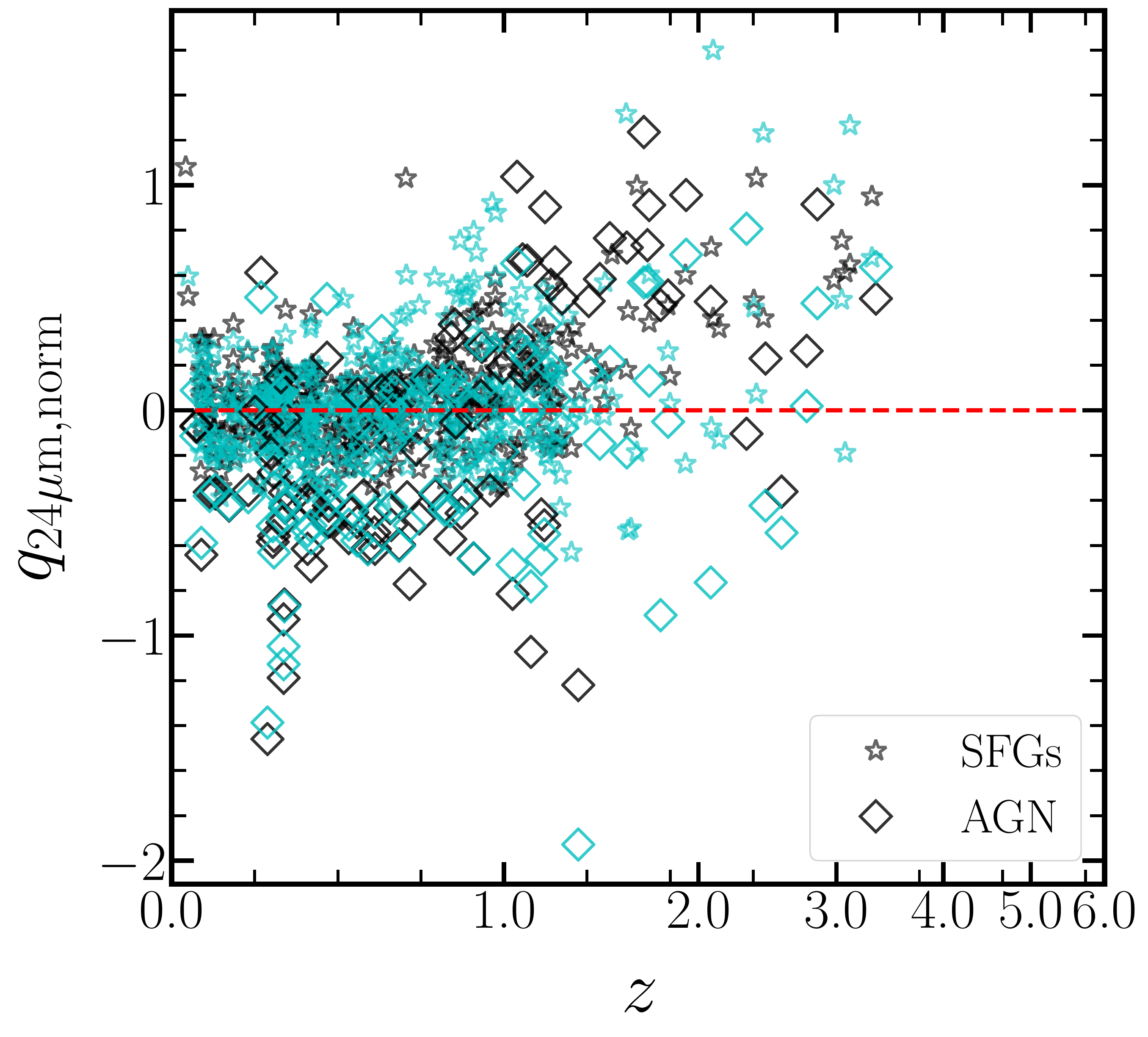}
    \caption{\textit{Left}: Variation of $k$-corrected $\qmir$ with redshift ($z$). The symbols are color-coded on the basis of their dust temperature ($\tdust$).
    \textit{Right}: Variation of normalised $\qmir$ ($q_{\rm 24\,\upmu m,norm}$) with $z$. The black symbols are for $\qmir$ in the observer's frame divided by the redshifted value of $\qmir$ from M\,82-type SED template, and the cyan symbols represents the $k$-corrected $\qmir$ using SED fitting divided by the mean value. The star and diamond symbols represent SFGs and AGN, respectively, in both the panels. }
    \label{fig:kcorrq24}
\end{figure*}

In the left panel of Fig.~\ref{fig:kcorrq24}, we show the variation of $\qmir$, computed using $k$-corrected emission at $24\,\upmu$m and at 1.4\,GHz, with $z$ for the 450 sources.\footnote{Note that, henceforth, we will present $k$-corrected $\qmir$ values unless mentioned otherwise. This is different from the $\qmir$ at observer's frame discussed in Section~\ref{sec:q24}.} The SFGs and AGN are shown as the star and diamond symbols respectively, and the colours represent their $\tdust$. For these sources, we find $\bra{\qmir} = 1.03 \pm 0.01$ with $1\,\sigma$ dispersion of 0.37. Hereafter, we will present the corresponding $1\,\sigma$ standard deviation in parenthesis. For comparison, we show the mean and $1\,\sigma$ dispersion of $\qmir$ from \citet{Appleton_2004} as the solid and dashed lines, respectively. At an average, we find $\bra{\qmir}$ for AGN to be slightly lower: $\bra{\qmir}_{\rm AGN} = 0.83\pm 0.01 (0.47)$ than that of SFGs: $\bra{\qmir}_{\rm SFG} = 1.09\pm 0.01 (0.32)$. However, within errors, this difference is insignificant. The values of $\qmir$ are listed in the top three rows of Table~\ref{tab:radioIR}.

The values of $\bra{\qmir}$ and its dispersion we have obtained for the sources in the ELAIS-N1 field are consistent with those reported in the literature for cosmologically distant sources. For example, \citet{Ibar_2008} find $\bra{\qmir} = 1.03 \pm 0.31$ which remains roughly constant in the range $0<z<1$, and up to $z\sim1.4$, \citet{Sargent_2010_1} find $\langle\qmir\rangle = 1.26 \pm 0.13$ for SFGs in the COSMOS field. However, on a cursory look, in contrast to previous studies, we find $\qmir$ to generally increase with $z$ in Fig.~\ref{fig:kcorrq24}, especially at $z\gtrsim 1.5$ for our radio selected sample of SFGs and AGN. This is a consequence of increasing $T_{\rm dust}$ with both $z$ and luminosity as seen in Fig.~\ref{fig:Tdust}, and can be gleaned from the left-hand panel of Fig.~\ref{fig:kcorrq24} which indicates that at a given redshift, the higher values of $\qmir$ correspond to higher $T_{\rm dust}$.

Often in the literature, $k$-correction at MIR wavelengths is performed, or $\qmir$ in the observer's frame is compared, by assuming a M\,82-like SED template \citep[e.g.,][]{Appleton_2004, Ibar_2008, Sargent_2010_1, Bonzini_2013, Ocran_2017}. Therefore, to compare our $k$-corrected $\qmir$ with that of a M\,82-like template, we compute the normalized $\qmir$ ($q_{\rm 24\upmu m, norm}$) for a source as the ratio of the observed $\qmir$ (points in Fig.~\ref{fig:q24}) to the mean value of $\qmir$ at the redshift of the source obtained from the M\,82 template, i.e., the solid black curve shown in Fig.~\ref{fig:q24}. In the right panel of Fig.~\ref{fig:kcorrq24}, we show the variation of $q_{\rm 24\upmu m, norm}$ from M\,82-like $k$-correction with $z$ in black, and of $k$-corrected $\qmir$ from our SED fitting, divided by $\bra{\qmir}$, in blue. From the figure, it is clear that both the methods indicate an increasing trend in $\qmir$ with $z$. Furthermore, it also indicates that $k$-correction of $\qmir$ obtained either by using a M\,82-like template, or by directly fitting the mid- to far-infrared SED using equation~\ref{eq:SED_IR}, do not show any strong systematic differences for the sample of radio selected sources from the uGMRT observations, apart from a constant offset of the absolute values of $\qmir$. The estimated $\qmir$ lies within 0.36\,dex of each other.

\subsubsection{$q_{\rm 70\upmu m}$} \label{sec:q70variation}

The variation of $\qfir$ with $z$ is shown in the left-hand panel of Fig.~\ref{fig:q70} where we have used the $k$-corrected flux densities at $70\, \upmu\rm m$ and 1.4\,GHz. The symbols and color scheme are the same as used in the left-hand panel of Fig.~\ref{fig:kcorrq24}. The solid black and the dashed lines show the mean $\qfir$ value of 2.15 and $1\,\sigma$ dispersion of 0.16 from \cite{Appleton_2004}. Table~\ref{tab:radioIR} tabulates the $\qfir$ values for both the classes of sources. The $\bra{\qfir} = 1.84 \pm 0.01 (0.42)$ for AGN is found to be significantly lower than the $\bra{\qfir} = 2.18 \pm 0.01 (0.26)$ for SFGs implying that, in the FIR wavelengths, an excess radio emission from AGN is better captured compared to $\qmir$. For the combined population, we find $\bra{\qfir} = 2.10 \pm 0.01 (0.34)$.

Our measured value of $\bra{\qfir}$ is close to that reported in the literature \citep[e.g.,][]{Appleton_2004, Sargent_2010_1, Mao_2011, Basu_2015}. Unlike $\qmir$ increasing with $z$, we find $\qfir$ to remain roughly constant up to $z = 1.5$. This is caused due to the increasing dust temperature with both, the redshift and the luminosity of the sources (see Fig.~\ref{fig:Tdust}).

Further, we have also determined the values of $\bra{\qmir}$ and $\bra{\qfir}$ for SFGs and AGN using rest-frame luminosity at 400\,MHz ($\lum$). These are listed in the Table \ref{tab:radioIR} (rows four--six). Overall, the mean $q$ obtained using $\lum$ are lower than that for $\Lum$ due to the nature of the radio continuum spectrum. We find the dispersion of $\qmir$ and $\qfir$ with respect to their respective mean values to be similar for both $\Lum$ and $\lum$ in SFGs, however, these are slightly higher for AGN at 400\,MHz.

\subsection{Variation of bolometric $q_{\rm TIR}$ parameter} \label{sec:qtirresults}

\subsubsection{Using monochromatic radio luminosities}

The statistical properties of monochromatic $q$, i.e., $\qmir$ and $\qfir$ are usually affected by the variation of $\tdust$ of the sources in a sample which gives rise to larger scatter, and/or systematic variation as seen in Fig.~\ref{fig:kcorrq24}. Therefore, total IR luminosity ($L_{\rm TIR}$) integrated between 8 and 1000\,$\upmu$m is used \citep[][]{Helou_1985, Bourne_2011}, and the corresponding $\qtir$ is defined as,
\begin{equation}
  q_{\rm TIR} = \log_{10}\left( \frac{L_{\rm TIR}}{3.75 \times 10^{12}~ {\rm [W]}}\right) - \log_{10}\left(\frac{\Lum}{[\whz]}\right).
     \label{eq_qIR}
\end{equation}

\begin{figure*}
\centering
    \includegraphics[height = 6.5cm]{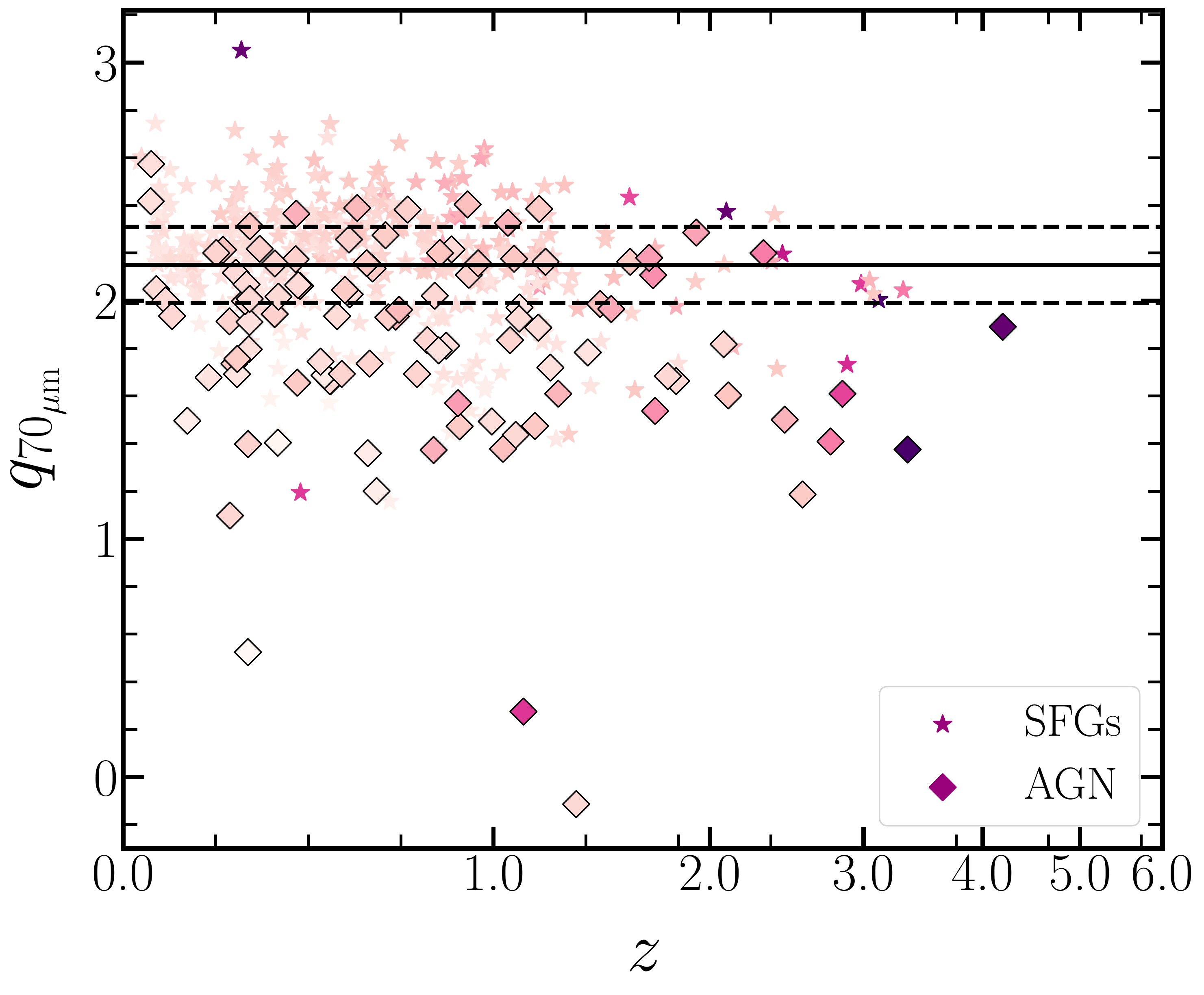}
    \includegraphics[height = 6.5cm]{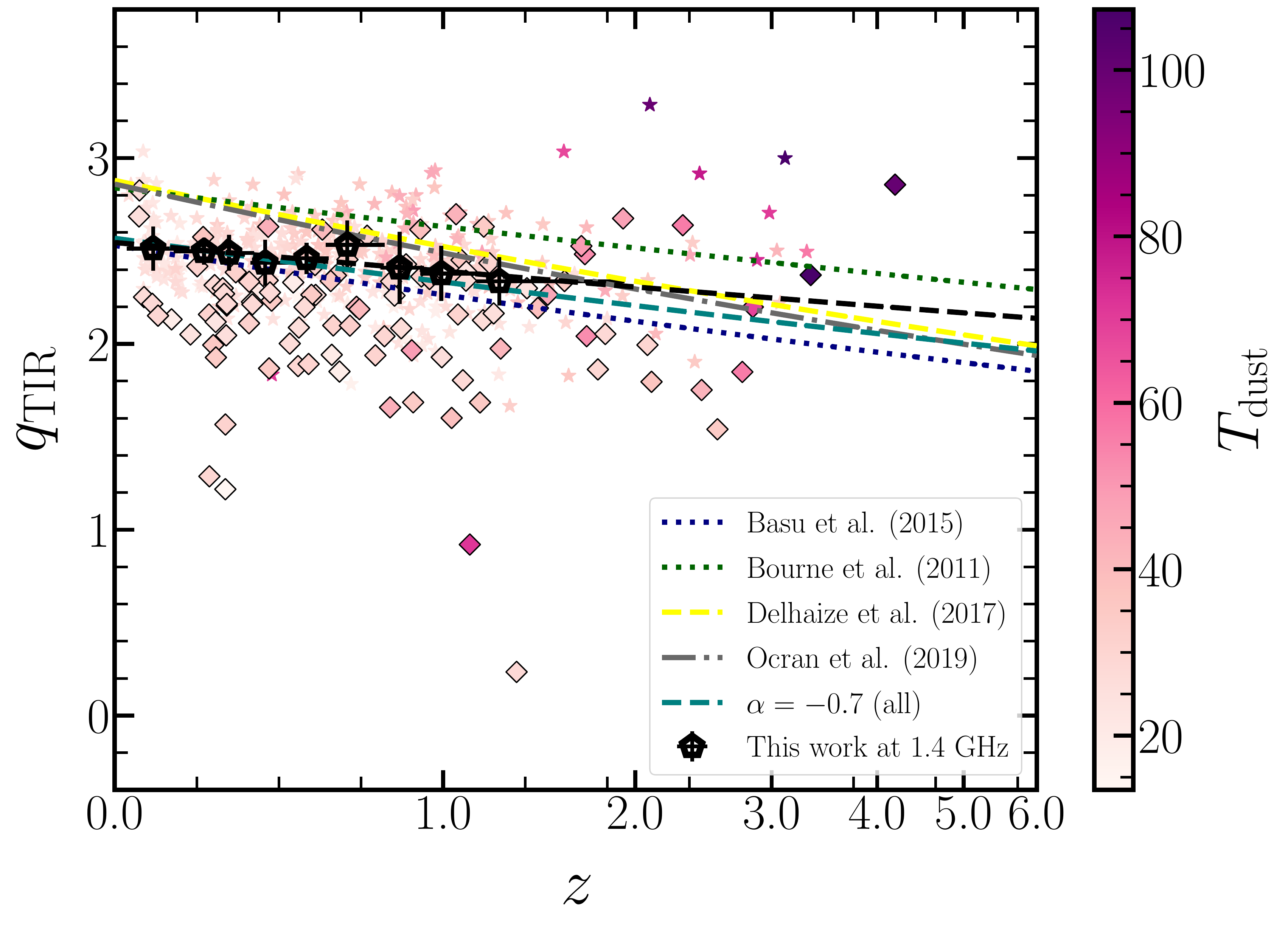}
    \caption{\textit{Left:} Variation of $k$-corrected $q_{\rm 70 \upmu m}$ values with redshift ($z$). The solid line represent the mean $\qfir$ value of $2.15$ with $1\sigma$ dispersion of 0.16 shown in dashed lines from \citet{Appleton_2004}. \textit{Right:} Variation of $\qtir$ with $z$. The black open pentagon denotes the median $\qtir$ values with each redshift bin for SFGs up to $z=2$ and the power-law fit for the same is shown in dashed black line. Other lines represent the evolution of $\qtir$ with $z$ from previous works reported in the literature. The star and diamond symbols represent the SFGs and AGN, respectively, in both the panels, and are colour coded based on dust temperature ($\tdust$).} 
    
    \label{fig:q70}
\end{figure*}

In the right panel of Fig.~\ref{fig:q70}, we show the variation of $\qtir$ with redshift. The SFGs and AGN are marked as star and diamond symbols, and their colours represent the $T_{\rm dust}$.
We measure a mean $\bra{\qtir}$ of $2.39\pm 0.01 (0.30)$ for the whole sample in the entire redshift range, which is in excellent agreement with previous estimates in the literature \citep[e.g.,][]{Bell_2003, Ivison_2010, Thomson_2014, Basu_2015, Ocran_2017}. The $\qtir$ value for our uGMRT sample remains roughly constant upto $z\sim1.5$ with  $\bra{\qtir}= 2.46\pm 0.01 (0.23)$ for SFGs across the full redshift range, whereas, AGN have a slightly lower $\bra{\qtir}= 2.15\pm 0.01 (0.38)$ and larger scatter compared to the SFGs. The values of $\qtir$ are tabulated in Table \ref{tab:radioIR}.

In the top panel of Fig.~\ref{fig:qTIR_hist}, the distribution of $\qtir$ for SFGs and AGN are respectively shown as the open and the filled histograms. From the distributions of $\qtir$ it is clear that for the SFGs, $\qtir$ has a symmetric distribution with smaller dispersion compared to AGN which shows a broader tail especially towards lower values of $\qtir$. This is a manifestation of the fact that there is an excess of radio emission in AGN as compared to that in the SFGs. We note that the tail towards lower values of $\qtir$ is likely to extend further. However, as discussed in Section~\ref{sec:agn_class}, the RL AGN population are missed in our sample due to the flux limits applied to our sample and perhaps obscuration at $24\,\upmu$m. Because of this, in our sample, a large fraction of the AGN are found to be overlapping with the SFGs in the radio--IR relations. This may indicate that the radio emission in these AGN, that are likely to be RQ\,AGN, is dominated by star formation.

\setlength{\tabcolsep}{4pt}
\renewcommand{\arraystretch}{.9} 
    \begin{sidewaystable}
    \footnotesize
        \centering
        \caption{Results of the $k$-corrected radio--IR relations for MIR emission at $\rm 24\,\upmu m$, FIR emission at $\rm 70\,\upmu m$, and total IR emission (TIR). $\bra{...}$ represents the mean values of $\qmir$, $\qfir$, and $\qtir$, along with their standard deviation ($1\,\sigma$) in parenthesis. The slopes $b$ are estimated by fitting a straight line to the radio vs. IR luminosities in the $\log\textrm{--}\log$ space. $\sigma_{\rm 24\upmu m}$, $\sigma_{\rm 70\upmu m}$ and $\sigma_{\rm TIR}$ are the 1\,$\sigma$ dispersion around the fit obtained by normalizing the $x$- and $y$-axes to their respective median values. The rows one--three are for rest-frame emission at 1.4\,GHz $k$-corrected using the $\alpha$ estimated in Section~\ref{sec:spec}; the rows four--six are for rest-frame emission at 400\,MHz $k$-corrected using the measured $\alpha$ similar to the top three rows; the rows seven--nine are for rest-frame emission at 1.4\,GHz $k$-corrected by assuming the same $\alpha=-0.7$ for all the sources; the bottom three rows are for the bolometric radio emission in the frequency range 0.1 to 2 GHz. }
        \begin{tabular}{lccccccccccccc}
            \hline
            \multicolumn{1}{c}{}  &
            \multicolumn{1}{c}{} &
            \multicolumn{4}{c}{$24\,\upmu$m}  &
            \multicolumn{4}{c}{$70\,\upmu$m}  &
             \multicolumn{4}{c}{TIR} \\
            \cline{3-6}
            \cline{7-10}
            \cline{11-14}
            
            Frequency & Type & $\langle q_{\rm 24 \upmu m}\rangle$ & $b$ & $\log a$ & $\sigma_{\rm 24\upmu m}$   & $\langle q_{\rm 70\upmu m}\rangle$ & $b$ & $\log a$ & $\sigma_{\rm 70\upmu m}$  & $\langle q_{\rm TIR}\rangle$ & $b$ & $\log a$ & $\sigma_{\rm TIR}$ \\
            \hline
            &&&&&&&&&&&&&\\
            1.4\,GHz & SFGs & $1.09 \pm 0.01 (0.32)$ & $0.93\pm0.02$ & $0.61\pm0.46$ &  1.45 & $2.18 \pm 0.01 (0.26)$ & $1.05\pm0.02$ & $-3.42\pm0.54$ & 1.10 &  $2.46 \pm 0.01 (0.23)$ & $1.07\pm0.02$ & $-4.34\pm 0.47$ & 0.81 \\ 
            & AGN & $0.83 \pm 0.01 (0.47)$ & $0.87\pm0.05$ & $2.36\pm1.14$ & 5.85 & $1.84 \pm 0.01 (0.42)$ & $1.08\pm0.07$ & $-3.11\pm1.67$ & 11.06& $2.15 \pm 0.01 (0.38)$ & $1.10\pm0.06$ & $-4.89\pm1.51$ & 6.89\\
            & SFGs+AGN & $1.01 \pm 0.01 (0.37)$ & $0.94\pm0.02$ &  $0.44\pm0.48$ & 4.51 & $2.10 \pm 0.01 (0.34)$ & $1.09\pm0.02$ & $-4.33\pm0.61$ & 7.88 & $2.39 \pm 0.01 (0.30)$ & $1.11\pm0.02$ & $-5.23\pm0.54$ & 7.28 \\
            &&&&&&&&&&&&&\\
\rule{0pt}{3ex}            400\,MHz & SFGs & $0.78 \pm 0.01 (0.32)$ & $0.93\pm0.02$ & $0.94\pm0.47$ & 1.15 & $1.87 \pm 0.01 (0.26)$ & $1.05\pm0.02$ & $-3.09\pm0.54$  & 1.13  & $2.15 \pm 0.01 (0.24)$ & $1.07\pm0.02$ & $-4.00\pm0.47$ & 0.81\\
            & AGN & $0.55 \pm 0.01 (0.47)$ & $0.90\pm0.05$ & $1.78\pm1.18$ & 10.44 & $1.56 \pm 0.01 (0.44)$ & $1.12\pm0.07$ & $-4.7\pm1.72$ &  10.54 &  $1.87 \pm 0.01 (0.40)$ & $1.15\pm0.06$ & $-5.73\pm1.54$ & 11.38\\
            & SFGs+AGN & $0.73 \pm 0.01 (0.38)$ & $0.94\pm0.02$ & $0.64\pm0.48$ & 6.95 & $1.80 \pm 0.01 (0.34)$ & $1.09\pm0.02$ & $-4.14\pm-0.60$ & 12.08 & $2.08 \pm 0.01 (0.30)$ & $1.11\pm0.02$ & $-5.04\pm0.53$ & 5.86\\
            &&&&&&&&&&&&&\\
\rule{0pt}{3ex}            1.4\,GHz & SFGs & $1.07 \pm 0.01 (0.31)$ & $0.95 \pm 0.02$ & $0.05\pm0.47$ & 1.85 & $2.15 \pm 0.01 (0.27)$ & $1.08 \pm 0.02$ & $-4.09\pm0.55$ & 1.09 & $2.43 \pm 0.01 (0.24)$ & $1.10 \pm 0.02$ & $-5.01\pm0.48$ & 0.70 \\
            ($\alpha = -0.7$) & AGN & $0.83 \pm 0.01 (0.46)$ & $0.90 \pm 0.05$ & $1.66\pm1.15$ & 6.06 & $1.84 \pm 0.01 (0.43)$ & $1.12 \pm 0.07$ & $-4.82\pm1.70$ & 9.64& $2.15 \pm 0.01 (0.38)$ & $1.14 \pm 0.06$ & $-5.77\pm1.50$ &  10.0 \\
            & SFGs+AGN & $1.01 \pm 0.01 (0.36)$ & $0.96 \pm 0.02$ & $-0.07\pm0.48$ & 5.26 & $2.08 \pm 0.01 (0.34)$ & $1.11 \pm 0.02$ & $-4.95\pm0.61$ & 11.05&  $2.37 \pm 0.01 (0.30)$ & $1.13 \pm 0.02$ & $-5.84\pm0.54$ & 5.53 \\
            &&&&&&&&&&&&&\\
\rule{0pt}{3ex}            Bolometric (RC) & SFGs & $0.59 \pm 0.01 (0.31)$ & $1.06\pm0.02$ & $-2.10\pm0.51$ & 1.38 & $1.68 \pm 0.01 (0.32)$ & $1.19\pm0.02$ & $-6.70\pm0.61$ & 1.03& $2.16 \pm 0.01 (0.23) $  & $1.07\pm0.02$ & $-3.85\pm0.41$ & 0.45 \\
            (0.1 to 2\,GHz) & AGN & $0.30 \pm 0.01 (0.45)$ & $1.02\pm0.05$ & $-0.71\pm1.23$ & 7.23 & $1.31 \pm 0.01 (0.49)$ & $1.26\pm0.07$ & $-8.08\pm1.85$ & 12.77& $1.87 \pm 0.01 (0.39)$ & $1.10\pm0.05$ & $-4.61\pm1.31$ & 6.62 \\
            & SFGs+AGN & $0.53 \pm 0.01 (0.37)$ & $1.07\pm0.02$ & $-2.33\pm0.52$ & 5.23 & $1.60 \pm 0.01 (0.39)$ & $1.24\pm0.03$ & $-7.78\pm0.67$ & 10.43  & $2.09 \pm 0.01 (0.30)$ & $1.10\pm0.02$ & $-4.61\pm0.46$ & 6.03 \\
            \hline
        \end{tabular}
        \label{tab:radioIR}
    \end{sidewaystable}

\subsubsection{Using bolometric radio luminosity} \label{sec:lrc}

Here we extend the radio--IR relation by using the bolometric radio continuum (RC) emission in our sample of sources. We computed the bolometric radio luminosity ($L_{\rm RC}$) for the 450\,sources by integrating their radio SEDs between the frequencies $\nu_1 =0.1\times 10^9$\,Hz and $\nu_2 = 2\times 10^9$\,Hz in the rest-frame as,
\begin{equation}
\frac{L_{\rm RC}}{[\rm W]} =  \int\limits_{\nu_1}^{\nu_2} \left(\frac{L_\nu}{[\whz]}\right) \,\left(\frac{{\rm d}\nu} {[\rm Hz]}\right).
\label{eq:Lrc}
\end{equation}
We note that, for a sample of nearby galaxies, \citet{Tabatabaei_2017} computed the radio emission integrated between 1 and 10\,GHz to study the distribution of $q$-parameters after separating the relatively high contribution of the thermal free--free emission in this frequency range.

To our knowledge, there is no standard definition of bolometric radio luminosity in the literature. Due to the relatively higher contribution of free--free emission above 2\,GHz \citep[e.g.,][]{Tabatabaei_2017, algera2021}, higher radio frequency emission would be contaminated. On the other hand, at frequencies below $\sim0.1$\,GHz, free--free absorption and/or ionization losses could affect the radio continuum emission in star forming galaxies \citep{Basu_2015b}. Both these effects tend to modify the power-law synchrotron spectrum. 
Therefore, we have avoided frequencies above 2\,GHz and below 0.1\,GHz for computing $\lrc$. Compared to $\Lum$ or $\lum$, $\lrc$ has the advantage of being dominated by synchrotron emission and is less susceptible to systematic and statistical fluctuations in the estimated values of $\alpha$. In Appendix~\ref{appendix}, we discuss the various advantages of using $\lrc$.

Using $L_{\rm RC}$, we define the different $q$-parameters as,
\begin{equation}
    q_{\lambda}^{\rm RC} = \log_{10}\left( \frac{L_{\lambda}}{[\whz] }\right) - \log_{10}\left( \frac{\lrc}{1.4 \times 10^9\, \rm{[W]} } \right),
    \label{eq:boloqtir}
\end{equation}
for monochromatic infrared luminosities ($L_{\lambda}$), where $\lambda = 24$ and $70\,\upmu$m, and,
\begin{equation}
    \qtirbol = \log_{10}\left( \frac{\ltir}{3.75 \times 10^{12} \,\rm{[W]} }\right) - \log_{10}\left( \frac{\lrc}{1.4 \times 10^9\, \rm{[W]} } \right),
    \label{eq:boloqtir}
\end{equation}
for total infrared luminosity. The $1.4\times10^9$ factor is for normalizing $L_{\rm RC}$ at 1.4\,GHz.

\begin{figure}
    \centering
    \begin{tabular}{c}
    \includegraphics[width=6cm]{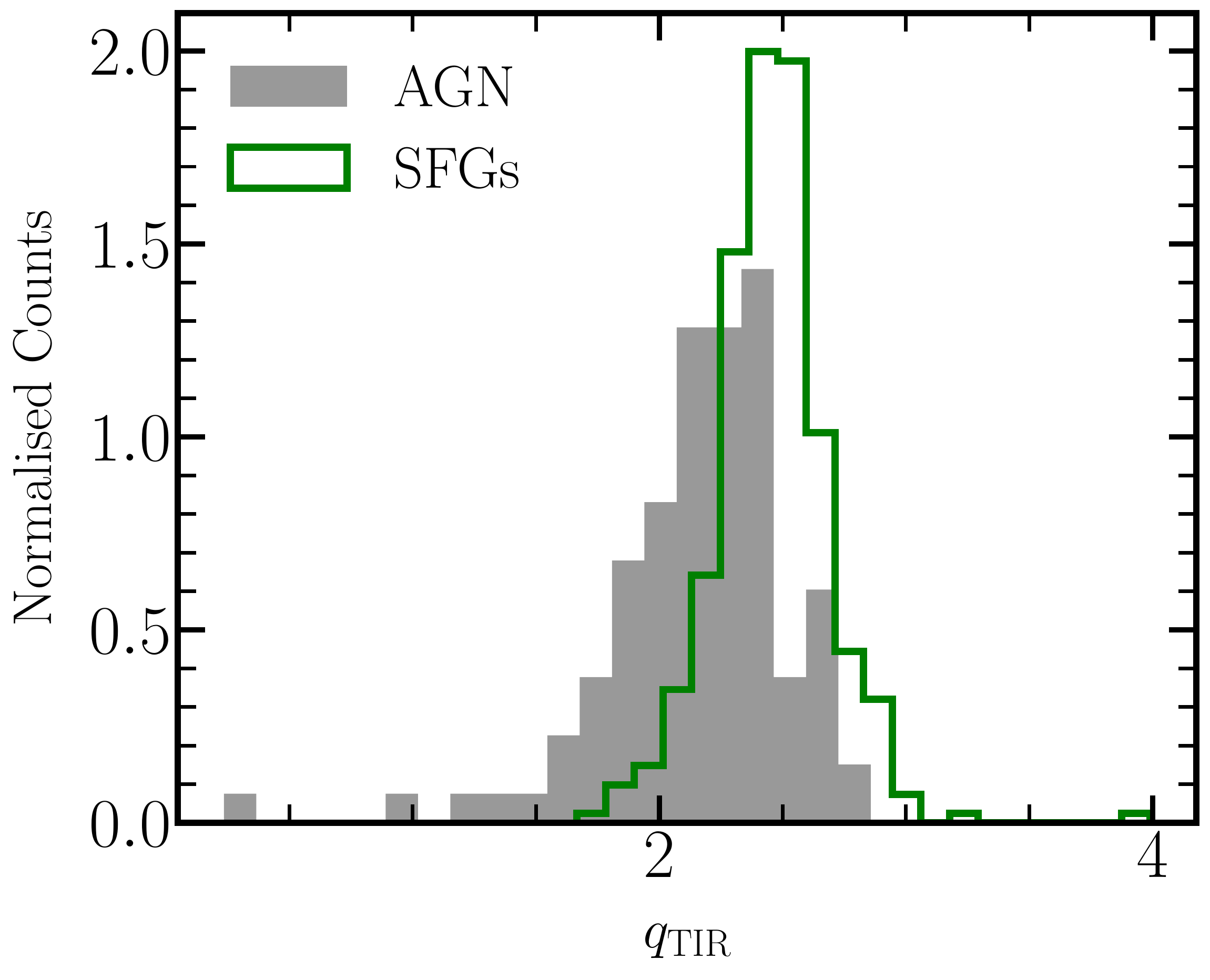}\\
    \includegraphics[width=6cm]{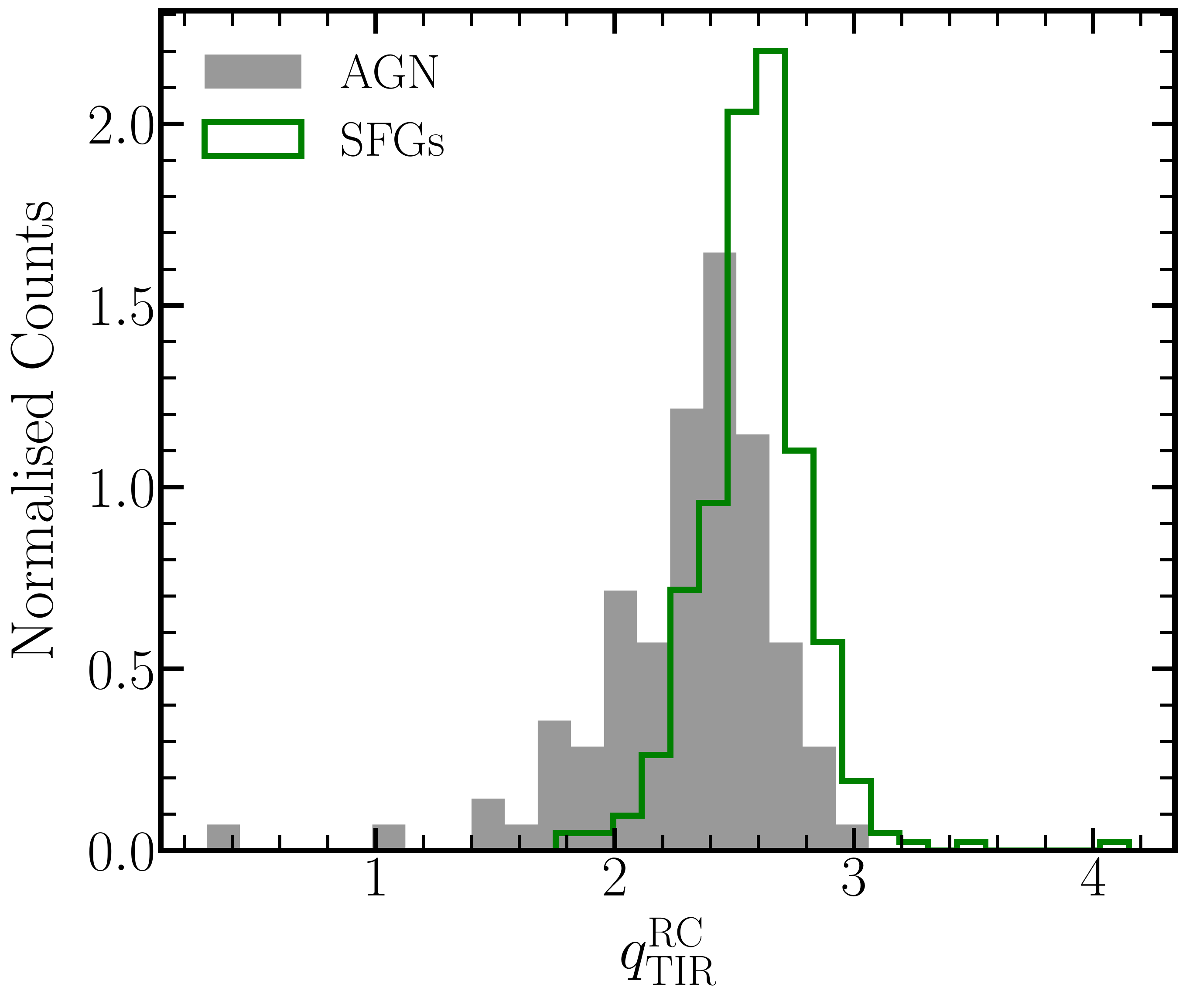}
    \end{tabular}
    \caption{Distributions of $\qtir$ (top panel) using monochromatic radio luminosity at 1.4\,GHz and of $\qtirbol$ (bottom panel) using bolometric radio luminosity. The open green and the shaded gray histograms are for SFGs and AGN, respectively.}
    \label{fig:qTIR_hist}
\end{figure}

The mean values of $q_{24\upmu \rm m}^{\rm RC}$, $q_{70\upmu \rm m}^{\rm RC}$ and $\qtirbol$ are listed in the bottom three rows of Table \ref{tab:radioIR}. 
The mean values of $\qtirbol$ for SFGs and AGN are found to be $2.16\pm 0.01 (0.23)$ and $1.87\pm 0.01 (0.39)$, respectively, and the AGN have larger relative scatter compared to the SFGs. For $\qtirbol$, the difference in the mean values for SFGs and AGN are more discernible compared to that for $\qtir$. This can be gleaned from the bottom panel of Fig.~\ref{fig:qTIR_hist} which shows indication that the peak of the distribution of $\qtirbol$ for AGN is shifted towards lower values compared to SFGs. Although there is significant overlap in the distribution of $\qtirbol$ between SFGs and AGN, $\lrc$ is perhaps a better measure to distinguish low luminosity AGN.

\subsection{Radio--infrared relations}\label{sec:irrc}

\subsubsection{Using monochromatic radio luminosity} \label{sec:monoradio-ir}

\begin{figure*}
    \centering
    \begin{tabular}{cc}
    \includegraphics[height=6.15cm,width=6cm]{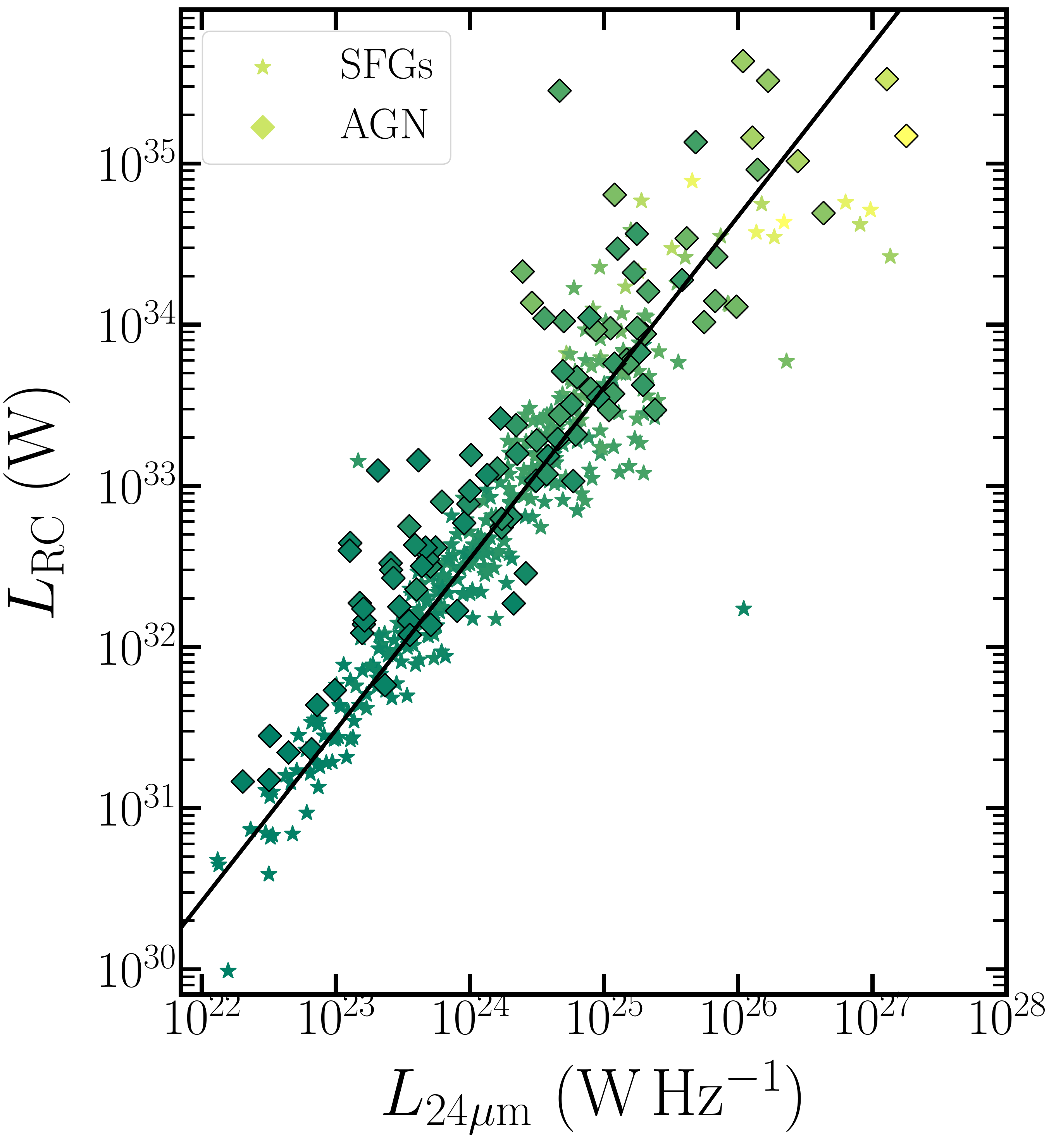} &
    \includegraphics[height=6.15cm,width=7cm]{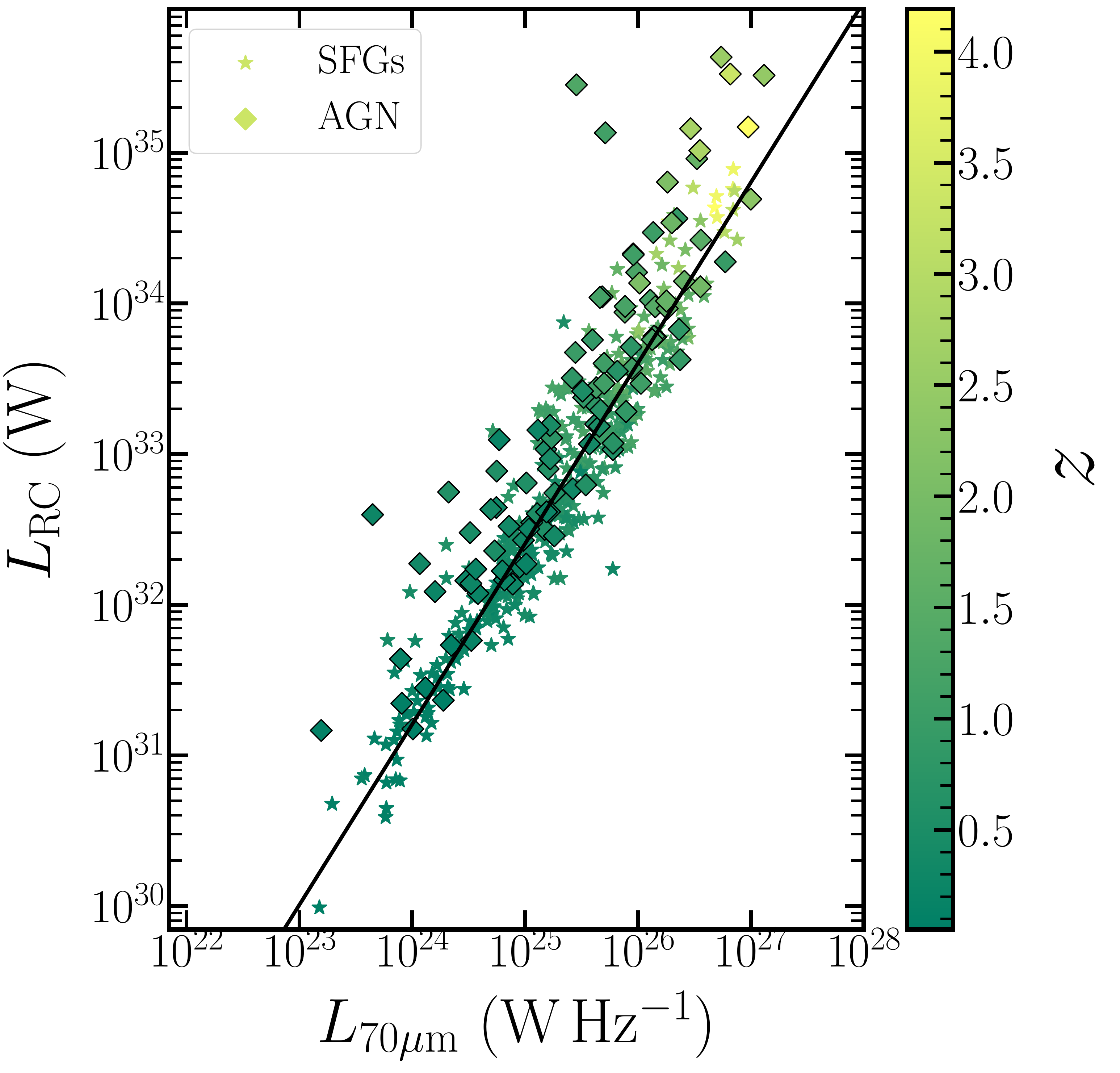} 
    \end{tabular}
    \caption{Variation of bolometric radio luminosity (from 0.1--2\,GHz) with rest-frame luminosity at $24\,\upmu$m ($L_{\rm 24\upmu m}$) in the left and at $70\,\upmu$m ($L_{\rm 70\upmu m}$) in the right.
    The star and the diamond symbols represents SFGs and AGN, respectively, and are coloured based on their redshifts. The solid lines in both the panels shows the best-fit straight line in $\log\textrm{--}\log$ space using only SFGs up to $z=2$. } 
    \label{fig:radio-IR}
\end{figure*}

Here, we study the variation of the radio luminosity $\Lum$ with different rest-frame infrared luminosities, namely, monochromatic $L_{\rm 24\upmu m}$ and $L_{\rm 70\upmu m}$ at $24\,\upmu$m and $70\,\upmu$m, and bolometric $L_{\rm TIR}$ of the total infrared emission. 
We fit these relations using orthogonal distance regression (ODR) in the $\log$--$\log$ space with the form $\Lum = a\,L_{\rm IR}^b$.
Here, $a$ is the normalization, $b$ is the slope (in $\log$--$\log$ space), and $\rm IR = 24\,\upmu m, 70\,\upmu m$ and TIR. Further, to avoid any biases that could arise from inadequate source identification and severe incompleteness in our flux-limited sample beyond $z \sim 2$, we have fitted using the data up to $z=2$. The values of $b$ obtained for the different radio--IR relations are listed in Table~\ref{tab:radioIR}. In the table, for completeness, we also present the slopes obtained for AGN, and for AGN and SFGs together.

For all the three types of radio--IR relations, we find the radio and infrared luminosities to be strongly correlated with Spearman's rank correlation $r > 0.9$. Of them, the correlation between $\Lum$ and $\ltir$ is found to be the strongest with $r = 0.95$. For our sample, we find the slope $b$ for all the relations to be non-linear with high statistical significance ($>3\,\sigma$). For SFGs, we find the $\Lum$--$\lmir$ relation to be sub-linear with slope $b = 0.93\pm0.02$, and the $\Lum$--$\lfir$ and $\Lum$--$\ltir$ relations to be super-linear with slopes $1.05\pm0.02$ and $1.07\pm0.02$, respectively. For the AGN, the radio--IR relations are slightly weaker compared to the SFGs with $r \sim 0.9$, and also have super-linear slopes of $1.10\pm0.06$ and $1.08\pm0.07$ for the $\Lum$--$\ltir$ and $\Lum$--$\lfir$ relations, respectively. 
Similar to SFGs, AGN also show a sub-linear slope of $b = 0.87\pm0.05$ for the $\Lum$--$\lmir$ relation. When both the SFGs and AGN are combined, the slopes for the three radio--IR relations do not change significantly (see Table~\ref{tab:radioIR}). This is perhaps due to the fact that MIR based sample-selection misses the radio bright AGN, while the RQ\,AGN in our sample follow the relations in a same way as the SFGs. Furthermore, within the uncertainties, the normalization $a$ of the relations are similar for both SFGs and AGN (see Table~\ref{tab:radioIR}). This further reiterates the fact that the radio emission in our sample of AGN, $\sim50$\,per\,cent of which are RQ\,AGN, are dominated by star-formation in the host galaxies as have been suggested previously \citep[e.g.,][]{Kimball_2009, Padovani_2011, Ocran_2017}.

\begin{figure}
    \centering
    \includegraphics[width = \columnwidth,height=7.25cm]{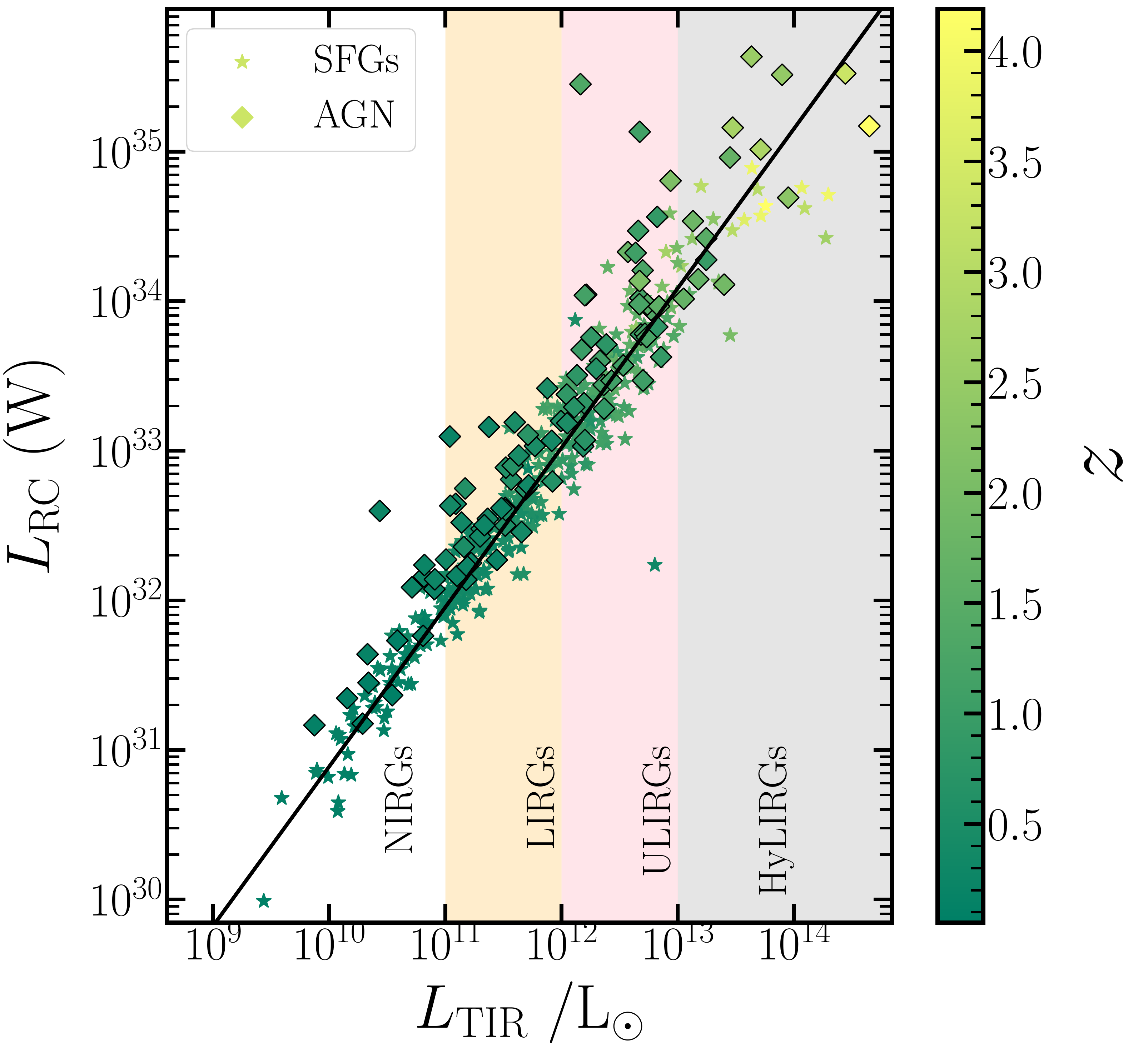}
    \caption{The variation of bolometric radio luminosity ($\lrc$) with total infrared luminosity ($\ltir$)). The best-fit straight line to these luminosities in $\log\textrm{--}\log$ for SGFs are shown by the solid black line up to $z = 2$. The star and diamond symbols represent SFGs and AGN, respectively, and are color coded on their redshifts. }
    \label{fig:lrcvsltir}
\end{figure}

The slope for both the $\Lum$--$\ltir$ and $\Lum$--$\lfir$ relations we obtained using our sample are in good agreement with  \citet{Bell_2003} who also reported a super-linear slope of $1.10 \pm 0.04$ for their sample of normal star-forming galaxies. A similar non-linear slope of $1.12 \pm 0.06$ was found by \cite{Basu_2015} by stacking blue-cloud galaxies up to $z = 1.2$ in the \textit{XMM}-LSS field. In fact, for a carefully selected sample of about 2000 SFGs at relatively low-$z$ ($<0.2$), a similar non-linear slope of $b = 1.11 \pm 0.01$ for the $\Lum$--$\ltir$ relation has been found \citep{Daniel_2021}. This suggests that the slope of the $\Lum$--$\lfir$ relation remains the same up to at least $z=2$ and that the non-linearity is likely to be intrinsic for SFGs irrespective of the selection criterion.

We have also studied the radio--IR relations at 400\,MHz using the rest-frame luminosity $\lum$. In rows four to six of Table~\ref{tab:radioIR}, we present the values of the slopes for all the three measures of infrared luminosity using $\lum$. The slopes of the $\lum$--$\lmir$, $\lum$--$\lfir$, and $\lum$--$\ltir$ relations are similar to those obtained using $\Lum$. We do not find any systematic differences in the statistical properties of the relations when $\lum$ is used instead of $\Lum$.

\subsubsection{Using constant $\alpha = -0.7$ for all sources}

In order to check whether $\Lum$ when computed using measured values of $\alpha$ has any systematic impact on the radio--IR relation as compared to when $\Lum$ is computed by assuming the same value of $\alpha$ for all the sources, we also studied the radio--IR relation using the latter. For this, we computed $\Lum$ by assuming $\alpha = -0.7$ for the same population of 450 sources and $k$-corrected the flux densities measured at 400\,MHz. The results of the radio--IR relations thus obtained are presented in Table~\ref{tab:radioIR} (rows seven--nine). Except for a slight decrease in the scatter of the radio--IR relations, we do not find any systematic differences in the mean values of the $q$ parameters and the slopes for this assumption on $\alpha$. This suggests that the method of $k$-correcting the radio flux densities does not systematically affect the radio--IR relations.

\subsubsection{Using bolometric radio luminosity} \label{sec:boloradio-ir}

We also study the variation of the bolometric radio luminosity integrated in the frequency range 0.1--2\,GHz with monochromatic infrared luminosities $\lmir$ and $\lfir$, and, total infrared luminosity $\ltir$ at high $z$. Fig.~\ref{fig:radio-IR} shows the variation $\lrc$ with $\lmir$ (left-hand panel) and with $\lfir$ (right-hand panel). In Fig.~\ref{fig:lrcvsltir}, we show the variation of $\lrc$ with $\ltir$. The solid lines show the best-fit using ODR as discussed in Sec.~\ref{sec:monoradio-ir}.

For our uGMRT sample, we find that all the three bolometric radio--IR relations, $\lrc$--$\lmir$, $\lrc$--$\lfir$ and, $\lrc$--$\ltir$ are strongly correlated with $r> 0.93$ (marginally stronger than the relations with $\Lum$) for SFGs and have super-linear slopes (rows ten--twelve in Table~\ref{tab:radioIR}).
The $\lrc$--$\ltir$ relation for SFGs show the strongest correlation with $r =0.96$ and slope $b = 1.07\pm0.02$. In fact, AGN also show a strong $\lrc$--$\ltir$ correlation with $r = 0.92$ (slightly weaker than SFGs) and a slope of $b = 1.10\pm0.06$. Interestingly, within the errors, the slopes of the $\lrc$--$\ltir$ relations for SFGs and AGN are found to be similar to that with monochromatic radio luminosities, however, the slopes are found to be significantly ($>4.5\,\sigma$) larger for $\lrc$--$\lmir$ and $\lrc$--$\lfir$ relations compared to the corresponding relations with $\Lum$ and $\lum$ for the SFGs (see Table~\ref{tab:radioIR}). For AGN, this increase in the slopes is at $\sim2\,\sigma$ level. Furthermore, in contrast to the sub-linear slopes we have found between monochromatic radio and $24\,\upmu$m luminosities, the slope is super-linear for $\lrc$. It is unclear what gives rise to this significant change in slopes of the bolometric radio--IR relations, and this trend requires to be investigated using multi-frequency radio continuum data in other well-known deep fields.

\section{Discussions}\label{sec:discussion}

Deep observations at radio frequencies in combination with multi-wavelength information, makes the statistical study of SFGs and AGN feasible. After broadly classifying the sources in the ELAIS-N1 field in our uGMRT observations at 400\,MHz into AGN and SFGs, we have studied the properties of the radio--IR relations. In this section, we discuss our results on these relations and their evolutionary properties in the context of SFGs in our sample. We will primarily focus on the properties of the radio--IR relation for monochromatic radio emission, and later compare them with what we observe for the bolometric radio emission in the context of their variation with redshift.

\subsection{Dispersion in \textit{q} parameters} \label{sec:dispersion}

In Section~\ref{sec:q}, we studied the variation of the monochromatic $q$ parameters at 24 and $\rm 70\,\upmu m$ with $z$, and in Section~\ref{sec:qtirresults} that of the TIR using the radio emission at 1.4\,GHz. In general, we find the relative dispersion\footnote{Defined as the ratio of $1\,\sigma$ standard deviation of the $q$ parameters to their corresponding mean values expressed in percentages.} of up to $\sim40$\,per cent in $\qmir$ to be significantly larger than those of $\qfir$ and $\qtir$ which have $\lesssim16$\,per\,cent dispersion for the whole sample. This behaviour is also true for the SFGs and AGN in our sample, however the dispersions of the $q$ values in SFGs are lower than the AGN (see Table~\ref{tab:radioIR}). To assess the impact on $q$ parameters for $k$-correcting the radio emission using the standard method of assuming a constant value of $\alpha$ for all the sources, we have also presented the values of $\bra{\qmir}$, $\bra{\qfir}$ and $\bra{\qtir}$ obtained by assuming a typical $\alpha = -0.7$ for all the sources in Table~\ref{tab:radioIR}. Within error, we do not find any significant difference in the statistical properties of the $q$ parameters when the radio emission is $k$-corrected by fitting the radio SED or by assuming a constant $\alpha$. Note that, from the radio SED fitting we find $\alpha$ to have a substantial scatter of $\sim26$\,per cent between sources (see right-hand panel of Fig.~\ref{fig:SED and alpha_dist_new}). This is expected to give rise to a larger scatter in the values of $q$ as compared to assuming a constant $\alpha$. Since both the $k$-correction methods yield similar scatter in $q$, it implies that the fluctuations in the physical parameters which determine the infrared luminosity, e.g., dust emissivity, $T_{\rm dust}$ and/or density of dust, and comparatively larger contamination from dust heated by AGN activity at MIR wavelengths introduces significant scatter in $q$ compared to fluctuations in $\alpha$ within the sample.

From the left-hand panel of Fig.~\ref{fig:kcorrq24}, we find the $\qmir$ values to increase with $z$ for our sample of sources in the ELAIS-N1 field, especially at $z \gtrsim 1$. Depending on the type of MIR SED template used for $k$-corrections, the value of $\qmir$ may vary strongly, especially towards higher redshifts \citep{Ibar_2008, Bourne_2011}.
Thus, in the right-hand panel of Fig.~\ref{fig:kcorrq24} we present the variation of $q_{\rm 24\upmu m,norm}$ with $z$ for 
$k$-corrections using our SED fitting discussed in Section~\ref{sec:IR_kcorr} and by using the M\,82-like SED template. We do not find any systematic difference between the two methods of MIR $k$-correction, demonstrating that $k$-correction using a M\,82-like template does not introduce any systematic variation in the values of $\qmir$ as compared to direct fitting of the mid- to far-infrared SED. This is perhaps not surprising because the high star-forming galaxy M\,82 in the nearby Universe is likely to be a prototypical example of the main sequence SFGs at high redshifts \citep{Magnelli_2009, Madau_2014, Bethermin_2017}. As $\qmir$ is relatively more sensitive to emission by hot-dust heated by the AGN activity, it shows the largest scatter among the three $q$ parameters (see Table~\ref{tab:radioIR}).

The trend of increasing $\qmir$ with $z$ seen in Fig.~\ref{fig:kcorrq24} is caused by the flux-limitation of our sample in the MIR where higher luminosity sources with higher $\tdust$ are preferentially being detected at higher redshifts. This can be gleaned from Fig.~\ref{fig:Tdust}. As the detected sources at $z\gtrsim1$ have higher $\tdust$, the peak of the dust emission shifts towards shorter wavelengths resulting in an increase of the monochromatic emission at $24\,\upmu$m. As a consequence $\qmir$ increases along with its dispersion. This brings out the critical fact about flux-limited study of the radio--IR relation at MIR wavelengths that $\tdust$ variation in the sample of sources introduces systematic and statistical biases. Furthermore, the PAHs are one of the major constituents of the interstellar dust that show broad emission features.
These PAH emission features at 7.7, 8.6, 11.3 and $12.7\,\upmu$m \citep[e.g.,][]{Roche_1991, Genzel_1998} in SFGs when redshifted beyond $z \approx 1$ falls in the $24\,\upmu$m band and therefore could also give rise to further scatter and biases.

In contrast to the $\qmir$, we observe $\qfir$ to remain nearly constant up to $z \sim 1$ and $\qtir$ shows a mild decrease with $z$. 
Both $\qfir$ and $\qtir$ have significantly lower dispersion compared to that of $\qmir$. This is because the dust emission near the peak of the infrared spectrum around $\sim70\textrm{--}80\,\upmu$m is mainly unaffected by the fluctuations in $\tdust$, and the 
total infrared emission is independent of $\tdust$. In previous studies, $\qfir$ has been observed to remain constant at $z \lesssim 1.5$ \citep[see, e.g.,][]{Sargent_2010_1, Basu_2015}, and is consistent with the observations of \citet{Smith_2014} who found the monochromatic $q$ values at wavelengths near the peak of the infrared emission to remain constant with $\tdust$. The increase in $T_{\rm dust}$ with both the luminosity and $z$ interplay in a way such that the decrease in monochromatic emission at $70\,\upmu$m due to the shift in the peak of the infrared spectrum towards shorter wavelengths is largely compensated by an increase of the radio luminosity towards higher redshift. This delicate balance results in $\qfir$ to remain roughly constant over a large redshift range.

Interestingly, we do not find any discernible trend with $\tdust$ in the variation of $\qtir$ as a function of $z$ in Fig.~\ref{fig:q70} (right-hand panel). This is because $\ltir$ integrated over the infrared spectrum is independent of $\tdust$. Thus, for the radio and MIR flux-limited sample used by us, $\qtir$ is a better indicator of the intrinsic evolution of the radio--IR relation. In Section~\ref{sec:qtir_evolution}, we will discuss about the mild decrease in $\qtir$ with $z$ in detail.

\begin{figure}
    \centering
    \includegraphics[width = 8cm]{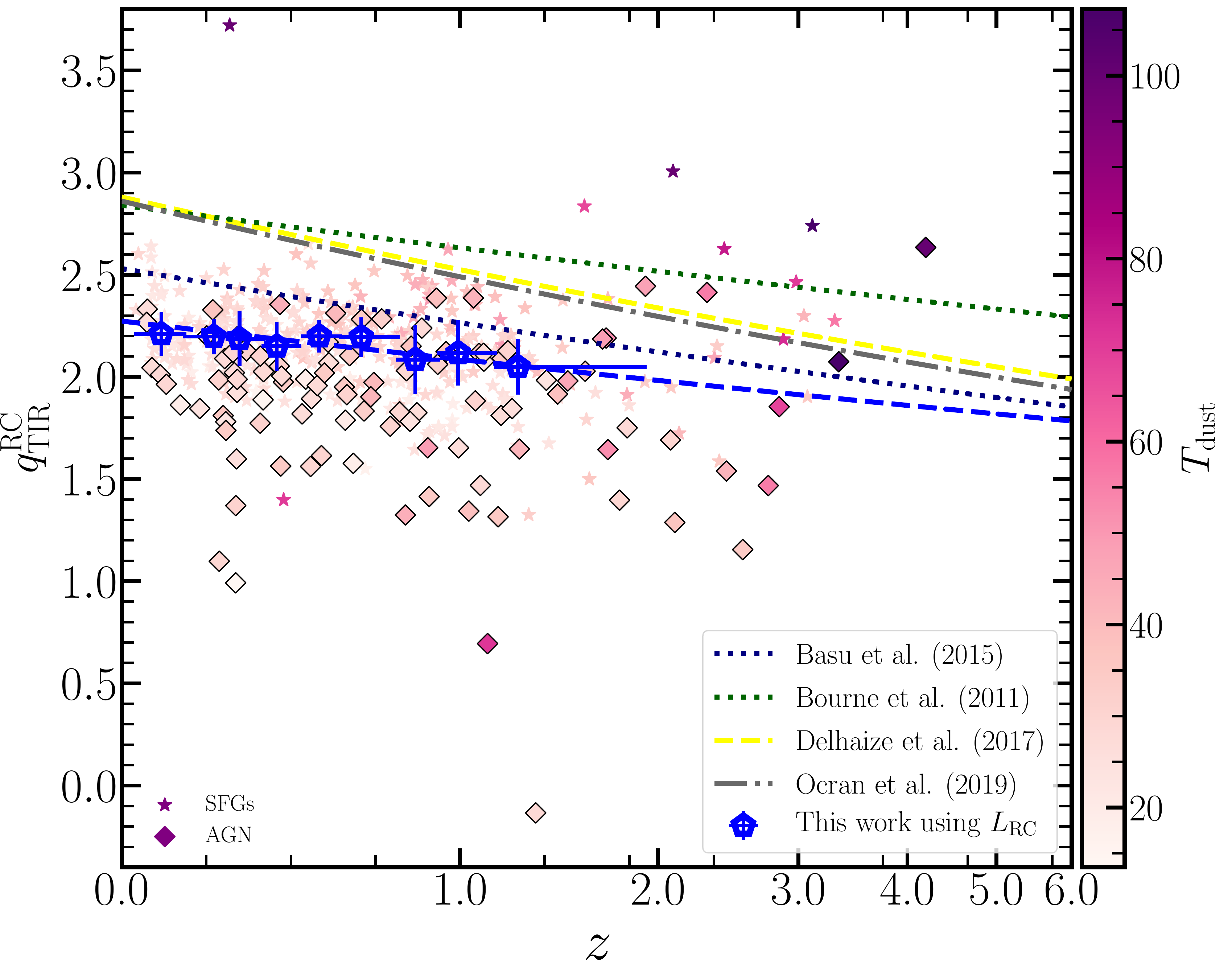}
    \caption{ \textit Variation of $\qtirbol$ with $z$. The blue open pentagon denotes the median $\qtirbol$ values with each redshift bin for SFGs up to $z=2$ and the power-law fit for the same is shown in dashed blue line. The error bars here represent the standard deviations in $\qtirbol$ values for each redshift bin. Other lines represent the evolution of $\qtir$ with $z$ from previous works reported in the literature. The star and diamond symbols represent the SFGs and AGN, respectively and are colour coded based on dust temperature ($\tdust$).} 
    \label{fig:qtirbol}
\end{figure}

\subsection{Scatter in the radio--infrared relations} \label{sec:scatter}

In Section~\ref{sec:irrc}, we studied the monochromatic radio--IR relations between $\Lum$ and $\lmir$, $\lfir$, and $\ltir$.
In our flux-limited sample of SFGs and AGN, we find the radio--IR relations, modelled as $\Lum = a\,L_{\mathrm{IR}}^b$, to be significantly non-linear (see Table~\ref{tab:radioIR}), wherein, for SFGs up to $z = 2$, the $\Lum$ \textit{versus} $\lmir$ relation is sub-linear with a slope $b = 0.93 \pm 0.02$; and $\Lum$ \textit{versus} $\lfir$, and, \textit{versus} $\ltir$ are super-linear with $b=1.05 \pm 0.02$, and, $1.07 \pm 0.02$, respectively. The sub-linear slope of the $\Lum$--$\lmir$ relation can be explained by the increase in luminosity at 24\,$\upmu$m with $\tdust$, that also manifests as an increase in $\qmir$ with $z$. In general, the slopes for AGN are larger than the SFGs, and for our sample as a whole, i.e., including both SFGs and AGN, the slopes of the radio--IR relations do not change significantly. This is due to the fact that, in deep radio observations, the sample is dominated by fainter SFGs, and contamination by AGN that are mostly RQ does not affect the radio--IR relations statistically. Except for the handful of radio luminosity-selected ($\Lum > 10^{25}\,\whz$) RL\,AGN in our sample (Section~\ref{sec:radio}), MIR- and spectroscopically-selected AGN mostly follow the radio--IR relations over the entire redshift range, up to $z \sim 4$, probed in this study. This makes it a challenge to identify AGN using the radio--IR relations alone. Nonetheless, this also implies that AGN contamination for a photometrically-selected sample of SFGs is unlikely to affect the estimation of the cosmic star formation history when the radio emission is calibrated using the radio--IR relations. This will form the basis of our forthcoming paper (Sinha\,A. et al., in preparation).

The non-linearity of the radio--IR relations implies that the $q$ parameters are expected to show a variation with redshift. For general expression of the radio--IR relations as $L_\nu = a\,L_{\mathrm{IR}}^b$, where $L_\nu$ is the luminosity at a radio frequency $\nu$, $q$ can also be written as,
\begin{equation}
\begin{split}
    q & = -\left( \frac{\log_{10}a}{b}\right) + \left(\frac{1-b}{b}\right)\,\log_{10}\,L_\nu\\
    & = -\left(\frac{\log_{10}a}{b}\right) + \left(\frac{1-b}{b}\right)\left[\log_{10}L_{\rm \nu, obs} - \alpha\,\log_{10}(1+z)\right].\\
\end{split}
\label{eq:q_with_lum}
\end{equation}
Here, $L_{\rm \nu, obs}$ is the radio luminosity in the observer's frame. This relation is true for both the monochromatic $\qmir$ and $q_{70\,\upmu \rm m}$, and for bolometric $q_{\rm TIR}$. It is clear that for non-linear slopes, the quantity $q$ depends on the radio luminosity, the slope, the spectral index, and the redshift. Hence, $q$ is expected to systematically vary with redshift. We therefore quantify the scatter ($\sigma_{\rm IR}$) directly from the radio--IR relation as, the $1\,\sigma$ dispersion of the quantity $\Delta$ defined as,
\begin{equation}
    \Delta = \left(\frac{L_\nu - a\,L_{\rm IR}^b}{a\,L_{\rm IR}^b}\right).
\end{equation}
The normalization $a$ and slope $b$ are obtained from fitting the corresponding radio--IR relation.

The scatter in the radio--IR relations for SFGs, AGN, and the combined population up to $z=2$ are listed in Table~\ref{tab:radioIR}. Firstly, the scatter of less than a factor of two for SFGs is significantly lower than the scatter of more than factor of $\sim3$ for AGN. Secondly, in general, the scatter in the $\Lum$--$\ltir$ relation of SFGs, $\stir = 0.81$, is lower when compared to the $\Lum$--$\lfir$/$\lmir$ relations ($\sfir = 1.10$/$\smir = 1.45$). This implies that the radio--IR relations are tighter when studied for the bolometric luminosity ($\ltir$) than the monochromatic $\lmir$ or $\lfir$ luminosities. This is expected because the infrared luminosity integrated over the infrared spectrum is independent of $\tdust$ while the monochromatic luminosities vary with $\tdust$. In Table~\ref{tab:radioIR}, we also present the scatter in the radio--IR relations measured by assuming the same $\alpha = -0.7$ for all the sources. Except for AGN, the slopes and scatter of the radio--IR relations for SFGs and the entire sample lies within the error when $\Lum$ is determined by assuming a constant $\alpha =-0.7$ or when $\alpha$ is measured for each source. However, the scatter $\stir = 0.70$ is slightly lower when measured using a constant $\alpha$. This was not readily evident from the scatter of $\qtir$ and $\qfir$. It is evident that for studying the intrinsic evolutionary properties of the radio--IR relations, the impact of the radio continuum spectrum needs to be investigated in detail when large surveys at multiple radio frequencies become available. 
This is because the nature of the radio continuum spectrum is determined by the mechanisms of cosmic ray particle injection and their subsequent energy loss by synchrotron and inverse-Compton cooling. These respectively depend on the star formation activity and magnetic field strengths in the SFGs. Otherwise, by assuming a constant $\alpha$, scatter introduced in the radio--IR relations due to fluctuations in the radio continuum spectra between sources are missed.

The scatter in the bolometric radio--TIR ($\lrc \textrm{--}\ltir$) relation for SFGs is significantly lower ($\sigma_{\rm TIR}^{\rm bol} = 0.45$) when compared to that for the monochromatic radio--TIR relation ($\sigma_{\rm TIR} = 0.81$; see Table~\ref{tab:radioIR}). This is because of the fact that $\lrc$ is largely insensitive to fluctuations of the synchrotron spectrum caused by energy loss/gain of the CREs. Interestingly, for monochromatic infrared emission, i.e., $\lrc$--$\lmir$ and $\lrc$--$\lfir$ relations, the scatter is comparable to those of $\Lum$--$\lmir$ and $\Lum$--$\lfir$ (also for $\lum$), suggesting that fluctuations in physical parameters that determine the monochromatic infrared emission, e.g., $\tdust$, dominates over fluctuations in synchrotron emission. An interesting feature of the bolometric radio--IR relations for AGN is that, although they have slopes and normalization similar to that of the SFGs (Sections~\ref{sec:monoradio-ir} and \ref{sec:boloradio-ir}), the scatter is significantly higher and remains unaffected for all the $\lrc$--$L_{\rm IR}$ relations.

As discussed before, it should be noted that, due to the flux limitations of the currently available near- and mid-infrared surveys, RL AGN populations are largely missing, and the so-called SFGs in our sample could be contaminated by LERGs, especially at redshifts above 0.5. It is interesting to notice that such contamination does not affect the radio--IR relations as all the non-AGN sources identified as SFGs in our study lie within the typical dispersion of the relations expected for SFGs. This suggests that the radio and far-infrared emission from mid-infrared selected sample of SFGs are likely to be dominated by the star formation activity.

\subsection{Apparent evolution of radio--IR relations with redshift}\label{sec:qtir_evolution}

The evolution of the radio--IR relation is typically studied by modelling the variation of the $\qtir$ with $z$ as $\qtir = q_0\,(1+z)^\upgamma$ \citep[e.g.,][]{Ivison_2010, Bourne_2011, Basu_2015, Calistro_2017, delvecchio2021}. Here, $q_0$ is the value at $z=0$ and $\upgamma$ is the exponent. A weak but significant decrease in $\qtir$ with $z$ has been observed in these studies. While it is clear from equation~\ref{eq:q_with_lum} that non-linearity of the radio--IR relations could in fact give rise to a variation in $q$ parameters with $z$, however, it remains unclear what gives rise to such a variation physically. 
In order to perform a similar modelling using our data, we binned the SFGs in our sample into nine redshift bins up to $z=2$ in a way that each bin contains an equal number of sources ($\sim40$ in our case) to avoid statistical biases introduced because of the binning, especially at higher $z$ where the number of sources are less. The median $\qtir$  values were computed for each of the redshift bins and was fitted using the form described above. These median $z$ and the corresponding median $\qtir$ values are listed in Table~\ref{tbl:qTIR_z_med}. 
We present the errors on the median $\qtir$ in each bin along with their $1\sigma$ dispersion in parenthesis. To reflect the robustness in the variation of $\qtir$ with $z$, we used the dispersions in each redshift bin as the errors while fitting.
Here, we will discuss about the variation of $\qtir$ because they are less prone to variations in $\tdust$ (see Section~\ref{sec:dispersion}). We find a mild variation of $\qtir$ with $z$ given by $ q_{\rm TIR} = (2.58 \pm 0.04)\ (1+z)^{-0.12 \pm 0.03}$, and is shown as the dashed black line in the right-hand panel of Fig.~\ref{fig:q70}. 
The index $\upgamma = -0.12\pm0.03$ for the SFGs in our sample is slightly 
lower than those reported previously. For example, \citet{Basu_2015} found $\upgamma = {-0.16 \pm 0.03}$ for star-forming blue-cloud galaxies in the \textit{XMM}-LSS field by extrapolating from data at 325\,MHz; \citet{Delhaize2017} found $\upgamma = {-0.19 \pm 0.01}$ for their sample in the COSMOS field; \citet{Calistro_2017} found $\upgamma = {-0.15\pm 0.03}$ for SFGs in the Bo\"otes field by extrapolating data from 150\,MHz; and \citet{Ocran2020} reported $\upgamma = {-0.20 \pm 0.02}$ for their sample of SFGs up to $z = 1.8$ in the ELAIS-N1 field.

A major difference of estimating $\qtir$ in these previous studies compared to ours is the way rest-frame luminosity at 1.4\,GHz is obtained. Most of the previous studies either used a constant spectral index value \citep[e.g.,][]{Basu_2015, Ocran2020}, or relied on spectral index measured between two frequencies with relatively shallower flux density cut-off compared to our study \citep[e.g.,][]{Calistro_2017, Delhaize2017}. Thus, to verify whether the measured spectral indices have any impact on the variation of $\qtir$ with $z$, we also estimated $\qtir$ values assuming a constant spectral index of $-0.7$ for our sample of SFGs. Following the same binning procedure described above, we obtain $q_{\rm TIR} = (2.57 \pm 0.03) (1+z)^{-0.14 \pm 0.02}$. This suggests that, for our sample, the method of $k$-correction plays little role in significantly affecting the variation of $\qtir$ with $z$.

In fact, for bolometric radio luminosity, although the dispersion of the $\lrc$--$\ltir$ is significantly lower, the relation remains super-linear, and as a consequence $\qtirbol$ is also observed to decrease with $z$. Similar to $\qtir$ estimated using $\Lum$, we find $\qtirbol$ to vary as $ \qtirbol = (2.27 \pm 0.03)\ (1+z)^{-0.12 \pm 0.03}$ and is shown as the blue dashed line in Fig.~\ref{fig:qtirbol}. This clearly indicates that the decrease in $\qtir$ or $\qtirbol$ with $z$ is an intrinsic feature of the radio--TIR relation.

Note that the variations of both $\qtir$ with $z$ depends on several physical factors that give rise to emission in the radio and infrared wavebands (see Section~\ref{sec:efficacy}), how they vary with $z$, and therefore in a way, on the sample selection. The slope of the radio--IR relations is also essential in determining the variation of $q$ with $z$ (see equation~\ref{eq:q_with_lum}).  For linear radio--IR relations, corresponding $q$ should remain constant with $z$. However, in Section~\ref{sec:irrc} we find the radio--IR relations to be significantly non-linear, and thus, the corresponding $q$ values are expected to vary with $z$. Following equation~\ref{eq:q_with_lum}, a super-linear slope for the $\Lum$--$\ltir$ and $\lrc - \ltir$ relations implies that, an increasing $\Lum$ and $\lrc$ with $z$ due to the usual Malmquist bias would naturally result in decreasing $\qtir$ and $\qtirbol$ with $z$.

\begin{table}
\centering
\caption{The median values of $\qtir$ and $\qtirbol$ of SFGs in different redshift bins. The standard deviation for each bin are shown in parentheses.}
\begin{tabular}{c c c c}
\hline 
Range of $z$ & Median $z$ & Median ${\qtir}$ & Median ${\qtirbol}$ \\[1ex]
\hline
0.030 -- 0.137 & 0.08 & 2.51 $\pm$ 0.01 (0.12) & 2.21 $\pm$ 0.01 (0.11) \\ 
0.137 -- 0.241 & 0.21 & 2.50 $\pm$ 0.01 (0.07) & 2.20 $\pm$ 0.01 (0.09) \\ 
0.242 -- 0.336 & 0.27 & 2.49 $\pm$ 0.01 (0.10) & 2.19 $\pm$ 0.01 (0.13) \\ 
0.337 -- 0.439 & 0.37 & 2.44 $\pm$ 0.01 (0.12) & 2.15 $\pm$ 0.01 (0.12) \\ 
0.448 -- 0.565 & 0.50 & 2.46 $\pm$ 0.01 (0.08) & 2.20 $\pm$ 0.01 (0.08) \\ 
0.567 -- 0.760 & 0.63 & 2.53 $\pm$ 0.01 (0.13) & 2.19 $\pm$ 0.01 (0.10) \\ 
0.760 -- 0.913 & 0.82 & 2.41 $\pm$ 0.01 (0.19) & 2.08 $\pm$ 0.01 (0.17)\\ 
0.914 -- 1.148 & 0.99 & 2.38 $\pm$ 0.01 (0.15) & 2.12 $\pm$ 0.01 (0.16) \\ 
1.154 -- 1.920 & 1.25 & 2.34 $\pm$ 0.01 (0.13) & 2.05 $\pm$ 0.01 (0.14)\\ 
\hline
\end{tabular}
\label{tbl:qTIR_z_med}
\end{table}

The slope $b$, however, is an essential parameter for studying the evolution of the radio--IR relations with $z$ as it is connected with various interdependent physical parameters of the ISM \citep[see e.g.,][]{Niklas_Beck_1997, Schleicher_Beck_2013, Basu_2015}. Therefore, it is crucial to pin down the origin of the non-linearity of the radio--IR relations in the context of SFGs at high redshifts. It is 
important to note that various observational selection effects may also result in an apparent evolution of $\qtirbol$ and $\qtir$ values as the faint sources may have been missed either because of the flux limitations at higher redshifts or due to obscuration \citep[also see][]{Daniel_2021}. As indicated by the different shaded bands in the right-hand panel of Fig.~\ref{fig:lrcvsltir}, the SFGs detected at higher redshifts ($z \gtrsim 0.5$) are dominated by luminous infrared galaxies (LIRGs), ULIRGs, and hyper-LIRGs (HyLIRGs). The star formation activity and magnetic field amplification in these galaxies often tend to be driven by mergers \citep{veill02, forst09, zhang10, kotar10, stott16, basu17a}. Although different galaxy-type are dominating the SFG population at different redshifts in our sample due to the evolution of the star-forming main sequence, it can be gleaned from Figs.~\ref{fig:radio-IR} and \ref{fig:lrcvsltir} that the radio--IR relations remain non-linear up to high luminosities with no obvious indication of a redshift evolution in the slope. It is possible that the heterogeneity of galaxy-type at high redshifts could give rise to the non-linearity in the radio--IR relations.

To mitigate this galaxy-selection bias, deep mid- to far-infrared observations are necessary to capture gas-rich, dynamically settled, star-forming galaxy populations at high redshifts. However, unfortunately, a super-\textit{Spitzer} or super-\textit{Herschel} space telescope is nowhere in the horizon. Therefore, spectroscopically confirmed normal galaxies up to a moderate redshift of $z \sim 0.5$ needs to be explored with deep radio observations using sensitive telescopes, such as the MeerKAT and upcoming SKA and ngVLA, combined with existing infrared surveys. For higher redshifts, stacking the existing infrared survey data at the location of the normal galaxies expected to be detected in deeper optical surveys with the JWST and the LSST is perhaps the only promising way forward to unravel the intrinsic redshift evolution of the radio--IR relations. The upcoming WHT Enhanced Area Velocity Explorer (WEAVE) survey\footnote{\url{https://www.ing.iac.es//confluence/display/WEAV/The+WEAVE+Project}} \citep{weave2014} will be important in advancing toward these directions.

\subsection{On the efficacy of using $\qtir$ and $\qtirbol$ to study ISM evolution} \label{sec:efficacy}

In a well selected sample of SFGs based on stellar mass and/or star formation activity, the $q$ parameter can perhaps be used to study the cosmic evolution of ISM. Besides equation~\ref{eq:q_with_lum}, assuming that a single-temperature dust emission is a good representation of the infrared SED, and the radio emission contains negligible contribution from the free--free emission, $q$ for monochromatic radio emission can also be expressed in terms of the physical properties of the ISM as \citep{Basu_2017},
\begin{equation}
    q_{\rm IR} = \log_{10}\,\left[\left(\frac{n_{\rm UV}}{n_{\rm CRE}} \right)\,\left(\frac{B_\lambda(\tdust)\,Q(\lambda,a)}{B_{\rm tot}^{1 - \alpha}\, \nu^\alpha} \right)\right] + C.
    \label{eq:qwithpars}
\end{equation}
Here, $n_{\rm UV}$ and $n_{\rm CRE}$ are the number densities of dust-heating UV photons and synchrotron emitting CREs; $B_{\rm \lambda}$ is the \textit{Planck} function; $Q(\lambda, a) \propto \lambda^\beta$ is the absorption coefficient for dust grains with radius $a$; $B_{\rm tot}$ is the total magnetic field strength; and $C$ is a normalization comprised of 
standard constants. 
All these parameters, namely, the ratio $n_{\rm UV}/n_{\rm CRE}$, $\tdust$, $B_{\rm tot}$, $\alpha$, and $Q$, can vary with redshift. Their interplay can therefore result in the variation of $q$ as a function of $z$. For galaxy-integrated emission, and for bolometric $\qtir$, $\tdust$ variation can be neglected, as indicated by our study. Therefore, by assuming negligible variation in dust properties with $z$, a decrease in $\qtir$ can be caused due to a combination of reasons, such as, flattening of the radio continuum spectrum, decrease in $n_{\rm UV}/n_{\rm CRE}$, and an increase in $B_{\rm tot}$ with $z$. It has been suggested in recent studies that the decrease in $\qtir$ is possibly caused due to increasing stellar mass at higher $z$ \citep{delvecchio2021} since at higher $z$, flux limited surveys are biased by more massive galaxies, or a consequence of selection bias based on star formation rate \citep{molnar2021}.

Our result on the variation of $\qtirbol$ brings out an important fact about the decrease in the values of $\qtir$. Note that, $\lrc$ is mostly immune to CRE energy gain/loss mechanisms which affect monochromatic radio emission at different frequencies differently.
As indicated by the data, $\qtir$ and $\qtirbol$ are largely independent of $\tdust$ (see Figs.~\ref{fig:q70} and \ref{fig:qtirbol}),\footnote{Since $\ltir$ is obtained by integrating over the dust emission, it is expected to be independent of $\tdust$. The variation seen in Fig.~\ref{fig:Tdust} is likely to be a consequence of flux limitation, especially above $200\,\upmu$m. For SFGs up to $z=2$, the sample of our interest, $\tdust$ and $\ltir$ are weakly correlated with $r=0.58$.} and $\lrc$ is independent of the radio continuum spectrum, equation~\ref{eq:qwithpars} simplifies as,\footnote{ Here, $\widetilde{C}$ is a different normalization constant compared to equation~\ref{eq:qwithpars}.}
\begin{equation}
 \qtirbol \approx  \log_{10}\,\left[\left(\frac{n_{\rm UV}}{n_{\rm CRE}} \right)\,\left(\frac{1}{B_{\rm tot}^{1 - \alpha}} \right)\right] + \widetilde{C}.  
 \label{eq:rctir}
\end{equation}

The flattening of radio continuum spectrum in high mass galaxies at high $z$ is unlikely to be the reason for a decrease in $\qtir$. Thus, if equation~\ref{eq:rctir} is a practical representation of $\qtirbol$, this implies that the evolution of magnetic fields with redshift and/or in different populations of SFGs plays an important role in shaping up the radio--IR relation. In the following, we discuss some of the plausible scenarios that can lead to a mild decrease in the values of $\qtirbol$ with $z$ from the perspective of the radio continuum emission.

(i) \textit{Evolving magnetic fields:} In order to produce a decrease in the values of $\qtir$ and $\qtirbol$ with redshift, equation~\ref{eq:q_with_lum} suggests that the radio luminosity should increase with redshift, and from equation~\ref{eq:rctir}, this can be caused due to an increase in magnetic field strengths with redshift. An increase of $B_{\rm tot}$ with redshift is possible because the small-scale ($\sim100$\,pc) turbulent dynamo action \citep[][]{Cho_2000, BS05, Gent_2013, Schleicher_Beck_2013, Schober_2016} generates stronger magnetic fields in high redshift massive galaxies that have higher gas \citep{Chowdhury_2020, Chowdhury_2021} and star-formation density \citep{Madau_2014, Pillepich_2018, Gruppioni_2020, Jo_2021}. The magnetic fields amplified by the action of 
turbulent dynamo can lead to a coupling between the magnetic fields and the gas densities, and therefore with star formation rate which is perhaps the cause of the non-linear radio--IR relations \citep{Niklas_Beck_1997, Schleicher_Beck_2013}. Furthermore, in addition to turbulence driven by star-formation, magnetic fields can also be amplified by galaxy merger-driven turbulence in the luminous galaxies at high redshifts \citep{veill02, Kilerci_2014, Whittingham_2021}. However, except for a handful of studies, a robust observational constraint on the evolution of magnetic fields with $z$ remains unclear \citep{oren95, berne08, joshi13, kim16, mao17}. On the other hand, since the mean-free path of dust-heating UV photons is $\mathcal{O}(50\,\rm pc)$, i.e., the size of the Str\"omgen sphere ionized by OB-type stars, UV photons are expected to remain trapped within the dense environment of massive galaxies at high $z$, except perhaps in Lyman-$\alpha$ emitters and in low metallicity galaxies near the epoch of reionization. That means, a decreasing $n_{\rm UV}/n_{\rm CRE}$ implies an increase in $n_{\rm CRE}$, which indicates an increased CRE injection at high $z$, likely due to the increase in cosmic star formation rate density up to $z\sim2$ \citep{Magnelli_2011,Madau_2014, Leslie_2020}. Additional data are required to pin down the cosmic evolution of magnetic fields and the escape fraction of UV photons in galaxies to unravel the cause of non-linearity in the radio--IR relations.

(ii) \textit{Evolving cosmic ray acceleration efficiency:} Another interesting possibility for the decrease of $\qtir$ with $z$ can be a super-linear dependence of
$n_{\rm CRE}$ with the star formation rate (SFR). To reproduce the variation of $\qtir$, $n_{\rm UV}/n_{\rm CRE}$ should vary with $z$ as $n_{\rm UV}/n_{\rm CRE} = n_0\,(1+z)^{-\beta} \propto (1+z)^\upgamma$, where $\beta > 0$. Since both $n_{\rm UV}$ and $n_{\rm CRE}$ are related to SFR, and $n_{\rm UV} \propto \rm SFR$, and say, $n_{\rm CRE} \propto \rm SFR^\delta$, then for $\qtir \propto (1+z)^\upgamma$ implies ${\rm SFR}^{1-\delta} \propto (1 + z)^\upgamma$. Thus, for a negative value of $\upgamma$, $\delta > 1$. This implies that at high redshifts, the acceleration efficiency of CREs in supernova remnants 
changes with SFR and/or there is a significant change in the initial mass function (IMF) where it becomes flatter having more massive stars. In order to establish the scenario of changing IMF, optical to infrared SED fitting in stellar-mass-selected galaxies in bins of redshift needs to be performed by including its variation. On the other hand, the scenario of changing CRE acceleration efficiency is not entirely unfeasible. Numerical simulations have suggested that the acceleration efficiency increases susbtantially with the Mach number ($\mathcal{M}$; \citealt{Caprioli2014, vanMarle2022}), where $\mathcal{M} = v_{\rm sh}/c_{\rm s} \propto \sqrt{\rho_{\rm gas}}$ for similar $v_{\rm sh}$. Here, $v_{\rm sh}$ is the shock velocity of supernovae explosions and $c_{\rm s}$ is the sound speed which depends on the gas density ($\rho_{\rm gas}$). Observations suggests that the average H{\sc i}-to-stellar mass ratio in star forming main-sequence galaxies increases from $\approx0.4$ in the local Universe to $\approx2.5$ at $z=1.3$ \citep{Saintonge2017, Chowdhury_2021}. This increase in the relative H{\sc i} mass hints at a possible increase in the gas density, and therefore an increase in the average $\mathcal{M}$ in high-$z$ galaxies which can give rise to an increased CRE acceleration efficiency and thus a super-linear dependence of $n_{\rm CRE}$ on SFR. Detailed semi-analytical calculations and/or numerical simulations are needed to quantitatively confirm this scenario.

These scenarios, however, can be further complicated depending on--- (i) whether or not energy equipartition between magnetic fields and CREs are valid \citep{Niklas_Beck_1997, Basu_2013, Basu_2017}; (ii) whether magnetic fields and gas densities are coupled; and (iii) whether high redshift SFGs are CRE calorimeters \citep[e.g.,][]{Werhahn_2021} or they lose CREs via winds on galactic scales \citep[e.g.,][]{Wiener_2017, Heald_2022}. In the first case, a breakdown in the energy equipartition due to CRE energy losses could also lead to an evolution of the radio--IR relation \citep[e.g.,][]{Schleicher_Beck_2013}. In the second case, magnetic fields amplified by the action of fluctuation dynamo can lead to a coupling between the magnetic fields and the gas densities, and therefore with the star formation rate. While in the third case, depending of the efficiency of cosmic ray escape from galaxies at high redshifts, the form of the radio--IR relations could be affected.
Hence, in order to infer the cosmic star formation rate density evolution at $z \gtrsim 2$ using radio continuum emission as a tracer via the radio--IR relations, the evolution of magnetic fields, gas densities and the rate of CRE escape also needs to be considered appropriately.

\section{Summary}\label{sec:summary}

We have performed deep observations of the ELAIS-N1 field using the uGMRT at 400\,MHz achieving a RMS noise of $15\,\upmu$Jy\,beam$^{-1}$ which provides $6\,\sigma$ point-source sensitivity of $\sim100\,\upmu$Jy\,beam$^{-1}$. A total of 2528 extragalactic sources were detected, of which 2321 sources having redshift information were broadly classified into SFGs and AGN using a host of publicly available multi-waveband data at infrared wavelengths and spectroscopy at optical wavelengths. About 24 and 76\,per cent of the sources were identified as AGN and SFGs, respectively, suggesting that at faint flux density end, SFGs dominate the population of extragalactic sources. Using these sources, we studied the statistical properties of the radio--IR relations, and our key findings are summarized below.
\begin{enumerate}[(i)]
    \item The median spectral index ($\alpha$) of the sources in the ELAIS-N1 field detected at 400\,MHz is found to be $\alpha = -0.58\pm0.15$. While, the median $\alpha$ for SFGs and AGN, when measured separately, are $-0.58\pm0.15$ and $-0.57\pm0.16$, respectively. Here the errors represent the median absolute deviation of the sample.
    \item The value of $\tdust$ for the sources in the ELAIS-N1 field is found to increase with $z$ and the total infrared (between $8\textrm{--}1000\,\upmu$m) luminosity ($\ltir$), implying that the three quantities are correlated, perhaps due to the flux limitations. As a result, the monochromatic luminosity at $24\,\upmu$m increases with $z$, which in turn results in $\qmir$ to increase with $z$. This implies that $\qmir$ is of limited value while investigating the radio--IR relations.
    \item The value of $\qfir$ mostly remain constant up to $z \sim 1$. Since the emission at $70\,\upmu$m lies near the peak of the infrared spectrum, $\qfir$ is less prone to variations in $\tdust$. However, $\qtir$ mildly decreases with $z$. Since the total infrared luminosity is independent of $\tdust$, $\tdust$ variation with $z$ does not affect the variation of $\qtir$. AGN shows lower $\qtir$ values than SFGs, signifying an excess radio emissions in AGN.
    \item We observe the $\Lum$--$\lfir$ and $\Lum$--$\ltir$ relations to have super-linear slopes while the $\Lum$--$\lmir$ relation is sub-linear for both SFGs and AGN.
    \item The statistical properties of $q$ and the slope of the radio--IR relations do not significantly depend on the method of how spectral indices are estimated using radio continuum observations between 0.1 and 1.4\,GHz for correcting the radio emission to rest-frame, i.e., directly from SED of each source or assuming the same value for all sources.
    \item For the first time, we investigated the radio--IR relations at high redshifts using the radio luminosity integrated between 0.1 and 2\,GHz ($\lrc$) that also exhibit super-linear slopes with various measures of infrared luminosity. The $\lrc$--$\ltir$ relation for SFGs is the tightest of all correlations with a scatter lower by a factor of $\approx2$ when compared to monochromatic $\Lum$--$\ltir$ and $\lum$--$\ltir$ relations. This is because, in contrast to monochromatic radio luminosities, $\lrc$ is independent of the fluctuations of the synchrotron spectrum caused due to CRE energy loss/gain mechanisms and/or contamination due to free--free emission in the sample of SFGs. 
    \item We present the variation of $q$ parameters derived using bolometric radio and IR luminosities as, $\qtirbol = (2.27 \pm 0.03)\ (1+z)^{-0.12 \pm 0.03}$. This moderate evolution could be attributed to the non-linearity of the $\lrc$--$\ltir$ relation, and suggests that non-linearity of the relations and variation of $q$ parameters have a common physical origin (see equation~\ref{eq:rctir}).
    \item The $q$ parameters depend on various physical parameters (see equation~\ref{eq:qwithpars}). From our results on the non-linear $\lrc$--$\ltir$ relation and decrease of $\qtirbol$ with $z$, we suggest that an increase in magnetic field strength and/or an increase in CRE acceleration efficiency with redshift are plausible reasons that could give rise to non-linearity in the radio--IR relations. More data are needed to investigate these scenarios.
\end{enumerate}



\section*{ACKNOWLEDGEMENTS}

We thank the anonymous referee for critical comments, and Dr Rainer Beck for insightful discussions. We thank E.\,F.\,Ocran for providing the M\,82 template for the $q_{\rm 24\upmu m}$ vs. $z$ variation. AS and AC would like to acknowledge DST for INSPIRE fellowship. AS would further like to thank Ramij Raja, Aishrila Mazumder and Sarvesh Mangla for fruitful discussions. AB thanks Dr G\"ulay G\"urkan and Dr Vijay Mahatma for helpful discussions on the properties of faint AGN in radio surveys. The authors thank the staff of the GMRT that made these observations possible. GMRT is run by the National Centre for Radio Astrophysics of the Tata Institute of Fundamental Research. 

This work is based in part on observations made with the Spitzer Space Telescope, which was operated by the Jet Propulsion Laboratory, California Institute of Technology under a contract with NASA.
This research has made use of data from HerMES project (http://hermes.sussex.ac.uk/). HerMES is a Herschel Key Programme utilising guaranteed time from the SPIRE instrument team, ESAC scientists and a mission scientist.
The HerMES data was accessed through the Herschel Database in Marseille (HeDaM - http://hedam.lam.fr) operated by CeSAM and hosted by the Laboratoire d'Astrophysique de Marseille.
HerMES DR3 was made possible through support of the Herschel Extragalactic Legacy Project, HELP (http://herschel.sussex.ac.uk).

Funding for SDSS-III has been provided by the Alfred P. Sloan Foundation, the Participating Institutions, the National Science Foundation, and the U.S. Department of Energy Office of Science. The SDSS-III web site is http://www.sdss3.org/.

SDSS-III is managed by the Astrophysical Research Consortium for the Participating Institutions of the SDSS-III Collaboration including the University of Arizona, the Brazilian Participation Group, Brookhaven National Laboratory, Carnegie Mellon University, University of Florida, the French Participation Group, the German Participation Group, Harvard University, the Instituto de Astrofisica de Canarias, the Michigan State/Notre Dame/JINA Participation Group, Johns Hopkins University, Lawrence Berkeley National Laboratory, Max Planck Institute for Astrophysics, Max Planck Institute for Extraterrestrial Physics, New Mexico State University, New York University, Ohio State University, Pennsylvania State University, University of Portsmouth, Princeton University, the Spanish Participation Group, University of Tokyo, University of Utah, Vanderbilt University, University of Virginia, University of Washington, and Yale University.

This research also made use of Astropy,\footnote{http://www.astropy.org} 
a community-developed core Python package for Astronomy \citep{astropy:2013, astropy:2018}, 
NumPy \citep{numpy11}, and Matplotlib \citep{matplotlib07}.

\section*{Data Availability}

The raw interferometric data from the uGMRT in Band\,3 are publicly available at the GMRT online archive (\url{https://naps.ncra.tifr.res.in/goa}) under project 32\_120.
The source classified catalogue at 400\,MHz including the spectral indices will be shared on a reasonable request to the corresponding authors.




\bibliographystyle{mnras}
\bibliography{references} 



\appendix

\section{ Comparison between bolometric and monochromatic radio luminosity}\label{appendix}

In Sec.~\ref{sec:lrc}, we have introduced the quantity $\lrc$, the integrated radio continuum luminosity in the range 0.1--2\,GHz. Here we discuss the various advantages of using $\lrc$ over $\Lum$.

\noindent{}(i) \textit{Dominated by synchrotron emission:} As the synchrotron emission typically has steeper spectrum ($\alpha \lesssim -0.5$) compared to the thermal free--free emission ($\alpha = -0.1$), it dominates at frequencies below $\sim2$\,GHz. For example, for a typical thermal fraction of $\sim10 (20)$\,per\,cent at 1.4\,GHz \citep{Basu_2013}, the contribution of free--free emission to $\lrc$ is $\lesssim7 (13)$\,per\,cent for synchrotron spectral index $\lesssim-0.5$.

\noindent{}(ii) \textit{Less susceptible to systematic variations of $\alpha$:}
Beside being dominated by synchrotron emission, using $\lrc$ has another advantage over $\Lum$ or $\lum$. The bolometric $\lrc$ is largely unaffected by possible systematic fluctuations of $\alpha$\footnote{Here we consider a systematic offset in the estimated values of $\alpha$.} in a sample of SFGs, either due to calibration offsets between data from different telescopes or due to flux measurement methods or due to $\alpha$ being affected by an increased contribution from free--free emission at the larger rest-frame frequencies. A large offset in the estimated value of $\alpha$ by up to $\sim 50$\,per cent (with respect to the true value of $\alpha$), affects $\lrc$ by less than $\sim 10$\,per\,cent. In contrast, the error in $\Lum$ or $\lum$ can be significantly more,\footnote{This depends on the $\alpha$ and slightly on the redshift of a source. Here the larger numbers correspond to a typical value of $\alpha$ in the range $-0.7$ to $-1$.} up to 40\,per\,cent, when $k$-corrected using the offset value of the measured $\alpha$.

\begin{figure}
    \centering
    
    \includegraphics[width=7.5cm] {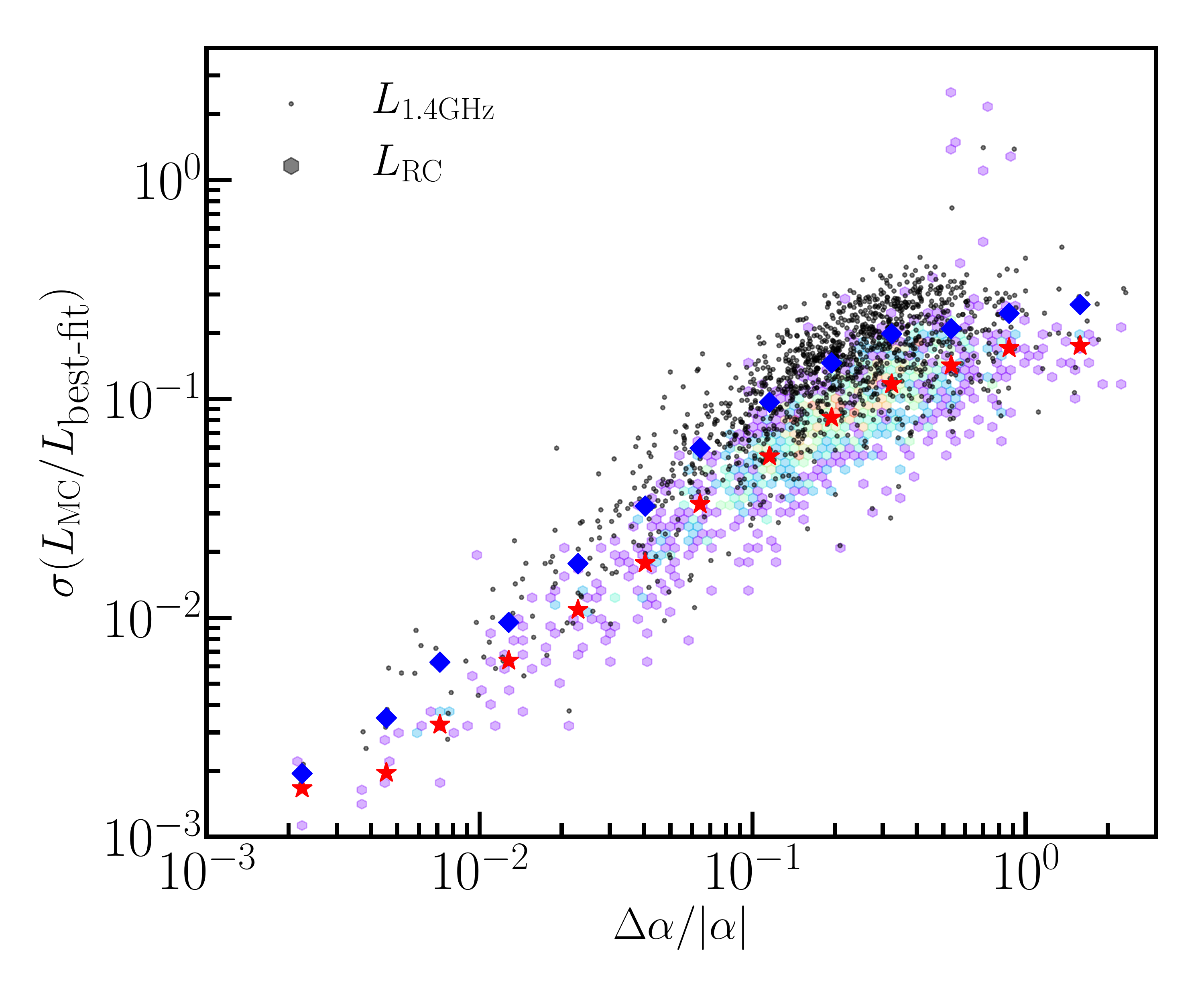}
    \caption{Variation of the dispersion $\sigma(L_{\rm MC}/L_{\textrm{best-fit}})$ for $\lrc$ (hexagons) and $\Lum$ (dots) as a function of the fractional uncertainty  values of the spectral index ($\Delta\alpha/|\alpha|$) for the 1278 sources in our sample for which we fitted the radio SED. $L_{\rm MC}$ is the luminosity obtained for each Monte-Carlo realization for a given source, and $L_{\textrm{best-fit}}$ is the best-fit luminosity. The red stars and the blue diamonds are the median values of the scatter in $\lrc$ and $\Lum$ in bins of $\Delta\alpha/|\alpha|$. The median fractional dispersion of $\Lum$ is larger by 20--90\,per\,cent compared to $\lrc$.}
    \label{fig:frac_uncer}
\end{figure}

\noindent{}(iii) \textit{Less susceptible to errors in $\alpha$:} $\lrc$ is mildly affected by statistical errors in the measured values of $\alpha$ and is therefore expected to capture the scatter in the radio--IR relations better compared to $\Lum$. To demonstrate the advantage of $\lrc$ over $\Lum$ for our sample, in Fig.~\ref{fig:frac_uncer} we show the variation of the standard deviation of the quantity $L_{\rm MC}/L_{\textrm{best-fit}}$, $\sigma(L_{\rm MC}/L_{\textrm{best-fit}})$, obtained from the Monte-Carlo simulations for each source discussed in Sec.~\ref{sec:spec}, as a function of the fractional error of $|\alpha|$, $\Delta\alpha/|\alpha|$, for $\Lum$ (black dots) and $\lrc$ (hexagons). For a source, $L_{\rm MC}$ is the corresponding luminosity for a Monte-Carlo realization, and $L_{\textrm{best-fit}}$ is the corresponding best-fit luminosity used in this work. It is clear that the scatter in the values of $\Lum$ is significantly larger than the scatter in the values of $\lrc$. The red stars and the black dots show the median values of $\sigma(L_{\rm MC}/L_{\textrm{best-fit}})$ in bins of $\Delta\alpha/|\alpha|$. In our sample, the scatter in $\Lum$ could be up to 90\,per\,cent larger than that of $\lrc$.

\bsp	
\label{lastpage}
\end{document}